\documentclass[aps,prc,showpacs,preprint,preprintnumbers,
showkeys,tighten,floatfix,nofootinbib]{revtex4}
\usepackage{amsmath,amssymb,amsfonts}
\usepackage{graphicx}
\usepackage{subfigure}
\usepackage{color}
\allowdisplaybreaks

\def \del{\partial}

\begin{document}

\title{Low Mass Dilepton Rate from the Deconfined Phase}
\author{Carsten \surname{Greiner}$^a$}
\author{Najmul \surname{Haque}$^b$}
\author{Munshi G. \surname{Mustafa}$^{a,b}$}
\author{Markus H. \surname{Thoma}$^c$}
\affiliation{$^a$Institut f\"ur Theoretische Physik, Johann Wolfgang Goethe 
University, Max-von-Laue-Strasse 1, D-60438 Frankfurt, Germany}
\affiliation{$^b$Theory Division, Saha Institute of Nuclear Physics,
1/AF Bidhannagar, Kolkata 700 064, India }
\affiliation{$^c$Max-Planck-Institut f\"ur extraterrestrische Physik, 
Giessenbachstrasse, 85748 Garching, Germany} 

\begin{abstract}
We discuss low mass dilepton rates ($\le 1$ GeV) from the deconfined phase
of QCD using both perturbative and non-perturbative models and compare with 
those from lattice gauge theory and in-medium hadron gas. Our analysis suggests 
that the rate at very low invariant mass ($ M\le 200$ MeV) using the 
nonperturbative gluon condensate in a semiempirical way within the Green 
function dominates 
over the Born-rate and independent of any uncertainty associated with the 
choice of the strong coupling in perturbation theory.  On the other hand the 
rate from $\rho-q$ interaction in the deconfined phase is important between 
$200$ MeV $\le M \le $ $1$ GeV as it is almost of same order of the 
Born-rate as well as in-medium hadron gas rate. Also the higher order 
perturbative rate, leaving aside its various uncertainties, from 
HTL approximation becomes reliable at $M\ge 200$ MeV and also becomes 
comparable with the Born-rate and the lattice-rate for $M\ge 500$ MeV, 
constraining on the broad resonance structures in the dilepton rate at 
large invariant mass. We also discuss the lattice constraints on the 
low mass dilepton rate. Furthermore, we discuss a more realistic way
to advocate the quark-hadron duality hypothesis based on the dilepton 
rates from QGP and hadron gas than it is done in the literature.
 
\end{abstract}

%\preprint{}
\maketitle

\section{Introduction}

The prime intention for ultra relativistic heavy-ion collisions is to study
the behaviour of nuclear or hadronic matter at extreme conditions like very
high temperatures and energy densities. A particular goal lies in the
identification of a new state of matter formed in such collisions, the
quark-gluon plasma (QGP), where the quarks and gluons are deliberated from 
the nucleons and move freely over an extended space-time region. Various
measurements taken in CERN-SPS~\cite{heinz} and 
BNL-RHIC~\cite{white,dilep,phot,ellip,jet,phenix} do lead to 'circumstantial
evidence' for the formation of QGP. Evidence is (or can only be) 
'circumstantial' because only indirect diagnostic probes exist. 

Electromagnetic probes, such as real photon and dileptons, are a 
particular example, and accordingly thermal dileptons have been 
theoretically proposed long time ago~\cite{larry}. At SPS
energies~\cite{agak} there was an indication for an enhancement of the dilepton 
production at low invariant mass ($0.2\le M(\mbox{GeV}) \le 0.8$ ) compared 
to all known sources of electromagnetic decay of the hadronic particles 
and the contribution of a radiating simple hadronic fireball (for
comprehensive reviews see Refs.~\cite{rapp1,rapp2,cass99}) . 
One of the possible explanations of this is the modification of 
the in-medium properties of the vector meson ({\em viz.,} $\rho$-meson) by
rescattering in a hadronic phase along with only the lowest order perturbative
rate, {\em i.e.}, $q\bar q$ annihilation from a
QGP~\cite{rapp1,rapp2,cass99,sum}. Also at RHIC energies~\cite{dilep}
a substantial amount of excess of electron pairs was reported in the
low invariant mass region. Models taking into account
in-medium properties of hadrons with various ingredients (see for
details~\cite{dz09,bcl09}) can not explain the data from RHIC in the
range $0.15 \le M (\mbox{GeV})\le 0.5$, whereas they fit the SPS data
more satisfactorily, indicating that a possible non-hadronic source
becomes important at RHIC.
 
On the other hand, the higher order perturbative calculations~\cite{agkz98} 
are also not very reliable at temperatures 
within the reach of the heavy-ion collisions. Moreover, perturbative 
calculations of the dilepton rate seem not to converge even in 
small coupling ($g$) limit.  Nevertheless, the lowest order perturbative 
$q\bar q$ annihilation is the only dilepton rate from the QGP phase that
is extensively used in the literatures. However, at large invariant
mass this contribution should be dominant but not at low invariant 
mass, where nonperturbative effects should play an important role. 
Unfortunately, the lattice data~\cite{karsch} due to its limitations 
also could not shed any light on the low mass dileptons.  However, the 
lattice calculations~\cite{boyd,lateos,peter} provide evidence for 
the existence of nonperturbative effects associated with the bulk 
properties of the deconfined phase, in and around the deconfined 
temperature, $T_c$. Also, indications have been found that the 
QGP at RHIC energies behaves more as a strongly coupled liquid 
than a weakly coupled gas~\cite{markus05}. Thus,  a nonperturbative 
analysis of the dilepton rate from the deconfined phase is essential. 

The dilepton emission at low invariant mass from the deconfined phase is 
still an unsettled issue in heavy-ion collisions at SPS and RHIC energies 
and, in particular, would be an important question for LHC energies and 
for compact baryonic matter formation in future FAIR energies~\cite{fair}, 
and also for the quark-hadron duality~\cite{rapp1,rapp2,qhdual} that 
entails a reminiscence to a simple perturbative lowest order $q\bar q$ 
annihilation rate~\cite{cley}. In this article we reconsider the dilepton 
production rates within the perturbative QCD, and non-perturbative models 
based on lattice inputs and phenomenological
$\rho - q$ interaction in the deconfined phase. The analysis suggests that 
the nonperturbative dilepton rates are indeed important at the low 
invariant mass regime. 

This article is organised in following way.
In sec. II we discuss the dilepton production rate from the deconfined phase
based on both perturbative and non-perturbative models. In sec. III
we compare the momentum integrated rates from both QGP and Hadron gas (HG).
We discuss the quark-hadron duality in sec. IV, and conclude in sec. V.

\section{Dilepton Rate From Deconfined Phase} 

The dilepton production rate can be derived from the imaginary part of
the photon self-energy~\cite{larry,gale} as

\begin{equation}
\frac{dR}{d^4xd^4P} = -\frac{\alpha}{12\pi^4} \frac{1}{e^{E/T}-1} \ 
\frac {\mbox{Im} \Pi_\mu^\mu(P)}{M^2} , \label{dilep_static}
\end{equation}
where $\alpha=e^2/4\pi$ and $P$ is four momentum of the virtual photon, 
$E$ is its energy, and we use
the notation $P\equiv (p_0=E,{\vec {\mathbf p}})$ and $p=|{\vec {\mathbf p}}|$. 
The square of the invariant mass of dilepton pair is $M^2=p_0^2-p^2$.  

\subsection{Born Rate }
To the lowest order the dilepton rate follows from one-loop photon self
energy containing bare quark propagators. This rate corresponds to
a dilepton production by the annihilation of bare quarks and antiquarks
of the QGP. Alternatively, this so-called Born-rate can also be obtained 
from the matrix element of the basic annihilation process folded with the
thermal distribution functions of quarks. In the case of massless lepton 
pairs in a QGP with two massless quark flavours with chemical potential 
one finds~\cite{cley} 
\begin{equation}
\frac{dR}{d^4xd^4P}= \frac{5\alpha^2}{36\pi^4} \frac{T}{p} \frac{1}{e^{E/T}-1}
\ln \frac{\left(x_2+\exp[-(E+\mu)/T]\right)\left(x_1+
\exp[-\mu/T]\right )}
{\left(x_1+\exp[-(E+\mu)/T]\right)\left(x_2+
\exp[-\mu/T]\right )} \ , \label{born_gen}
\end{equation}
where $x_1=\exp[-(E+p)/2T]$, $x_2=\exp[-(E-p)/2T]$. A finite quark mass 
can easily be included.

For $\mu=0$ the dilepton rate becomes
\begin{equation}
\frac{dR}{d^4xd^4P}= \frac{5\alpha^2}{18\pi^4} \frac{T}{p} \frac{1}{e^{E/T}-1}
\ln \left (\frac{\cosh\frac{E+p}{4T}}{\cosh\frac{E-p}{4T}}\right) \ , 
\label{rate_born}
\end{equation}
whereas that for total three momentum ${\vec {\mathbf p}}=0$ is given as
\begin{equation}
\frac{dR}{d^4xd^4P}= \frac{5\alpha^2}{36\pi^4}\ n(E/2-\mu) \ n(E/2+\mu)
\ , \label{born_mup0}
\end{equation}
with $n(y)=(\exp(y)+1)^{-1}$, the Fermi-Dirac distribution function.

\subsection{ Hard Thermal Loop perturbation theory (HTLpt) Rate }
In order to judge the reliability of the lowest order result, one should 
consider higher order corrections. These corrections involve quarks and gluons 
in the photon self energy beyond the one-loop approximation. Using bare 
propagators at finite temperature, however, one encounters infrared 
singularities and gauge dependent results. These problems can be resolved, at
least partially, by adopting the Hard-Thermal Loop (HTL) resummation 
scheme~\cite{braaten}.
The key point of this method is the distinction between the soft momentum
scale ($\sim gT$) and the hard one ($\sim T$), which is possible in the weak
coupling limit ($g<<1$). Resumming one-loop self energies, in which the loop
momenta are hard (HTL approximation), effective propagators and vertices 
are constructed, which are as important as bare propagators if the momentum 
of the quark or gluon is soft. In HTLpt the bare $N$-point functions (propagator and 
vertices) are replaced by those effective $N$-point HTL functions which 
describe medium effects in the QGP such as the thermal masses for quarks 
and gluons and Landau damping.

The importance of the medium and other higher order effects on the dilepton
rate depends crucially on the invariant mass and the momenta of the virtual
photon. Therefore, we will discuss now the different kinematical regimes:

\begin{figure}[!tbh]
\vspace*{0.40in}
\subfigure{
\includegraphics[height=0.42\textwidth, width=0.45\textwidth]{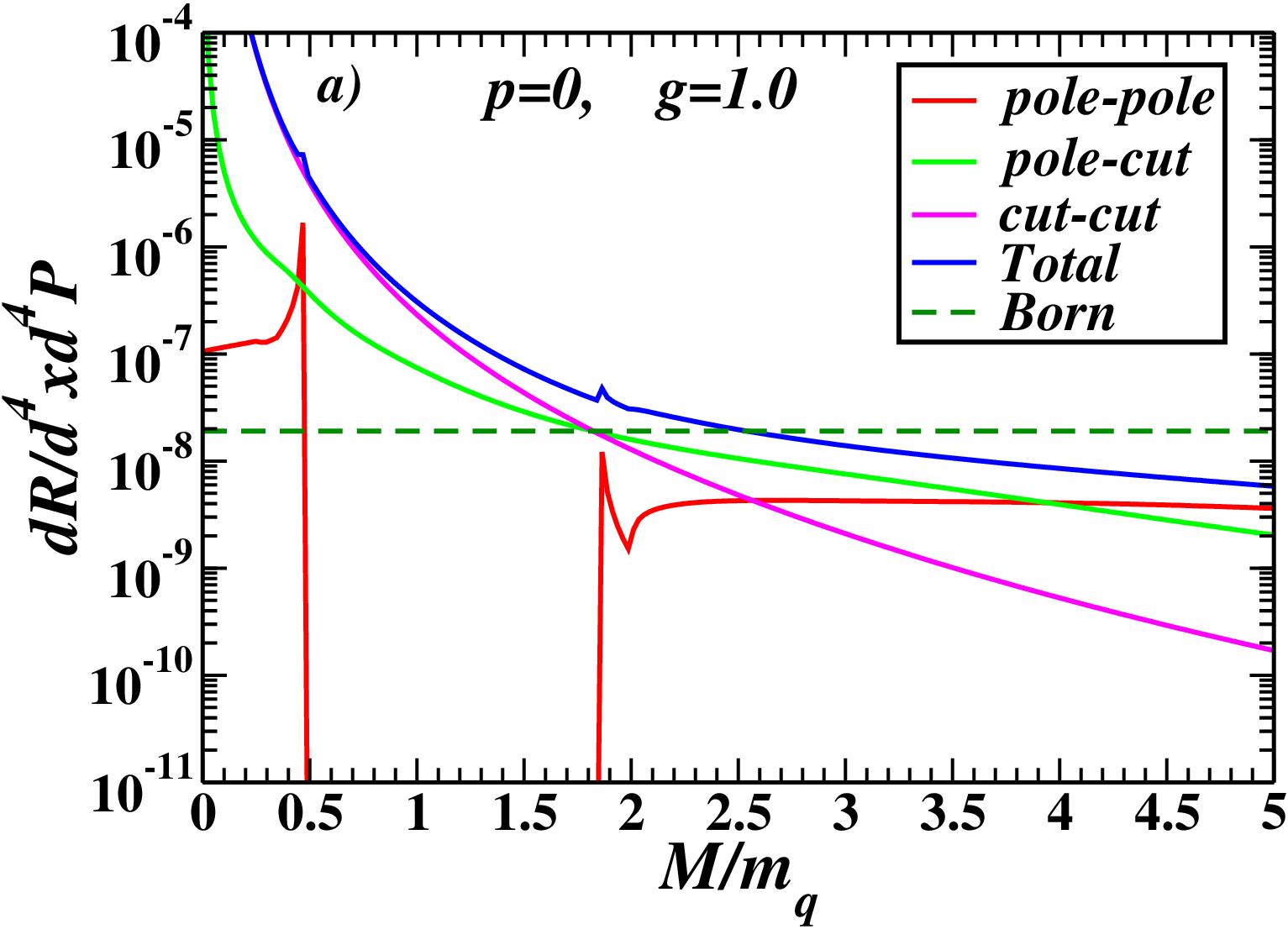}}
\subfigure{
\includegraphics[height=0.42\textwidth, width=0.45\textwidth]{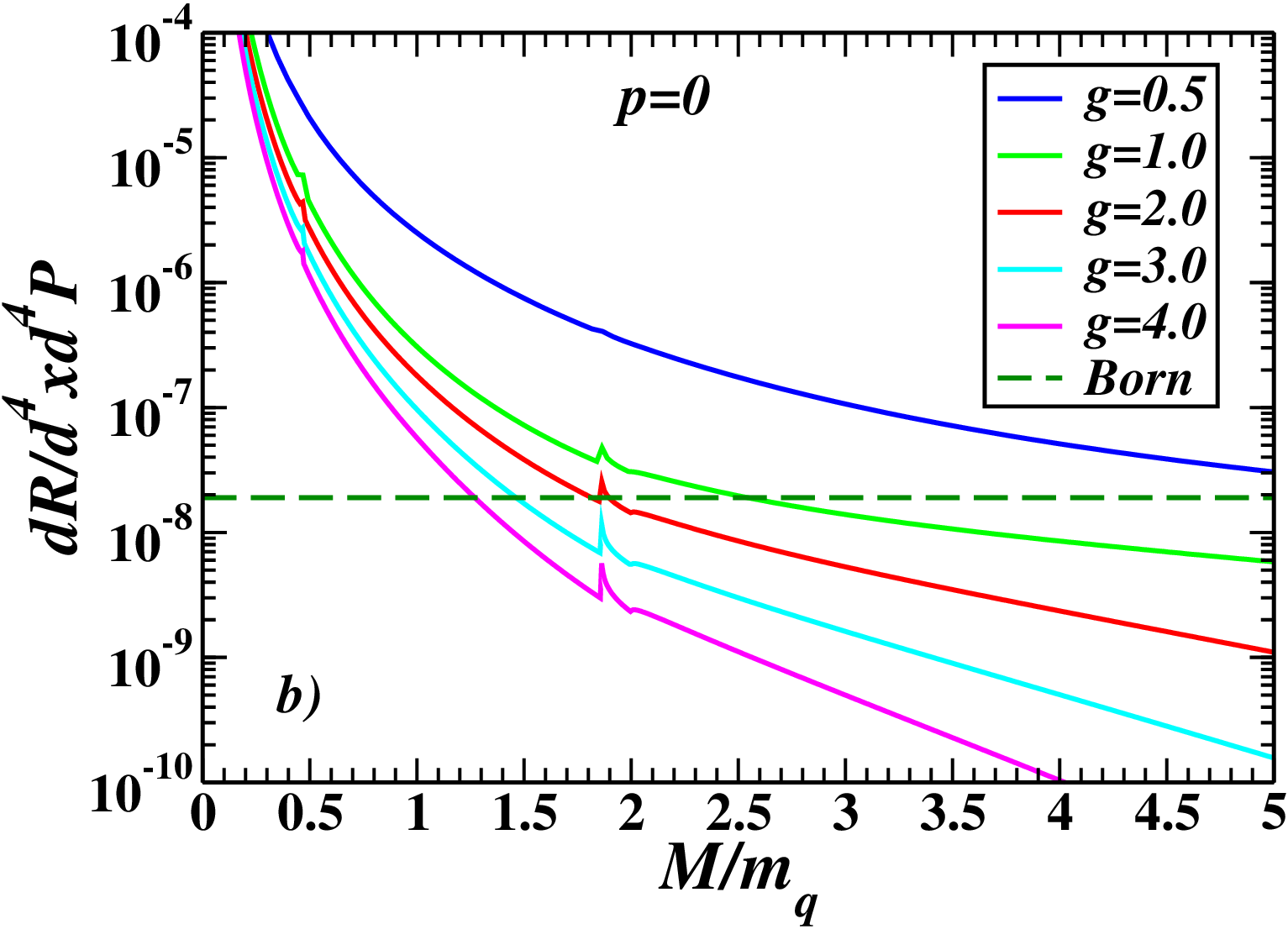}}
\vspace*{-0.08in}
\caption{(Color online){\em Left panel (a):} 1-loop dilepton rate for small
 invariant masses $M\sim gT$ at zero momentum and Born-rate (dashed line) 
versus the scaled invariant photon mass $M/m_q$ for $g=1$. The van Hove peaks 
and energy gap are evident in the 1-loop rate. {\em Right panel (b):} Total
1-loop rate for various $g$ values.
}
\label{fig_1loop}
\end{figure}

\subsubsection{Soft Rate ($M\sim gT$ and $p\sim gT$)}
For soft invariant masses\footnote{Note that for ultrasoft $M\sim g^2T$ and
arbitrary momentum the rate is non-perturbative and cannot be calculated even 
within the HTL improved perturbation theory. This observation holds in 
particular for real hard photon~\cite{agz00}.} and momenta of order $gT$ 
one has to use HTL quark propagators and vertices in the one-loop photon 
self energy. These corrections are of same order as the Born-term~\cite{yaun}. 
Physically these corrections correspond to two different processes. First 
the poles of the HTL resummed quark
propagators describe quasiparticles in the QGP with an effective thermal 
quark mass of the order of $gT$. Hence dileptons are generated by the 
annihilation of collective quark modes instead of bare quarks. In particular
the HTL quark dispersion contains a so called plasmino branch which exhibits
a minimum at finite momentum. This nontrivial dispersion leads to sharp
structures (van Hove singularities and energy gap) in the dilepton production 
rate\footnote{For a discussion of van Hove singularities in the QGP at 
${\vec {\mathbf p}}=0$ see Refs.~\cite{yaun,markus,munshi} and also 
Ref.~\cite{wong} for 
${\vec {\mathbf p}}\ne 0$.} in contrast to smooth Born-rate.
Secondly, the imaginary part of the HTL quark self energy containing 
effective HTL $N$-point (propagators and quark-photon vertex) functions 
corresponds to processes involving the absorption or emission of thermal 
gluons. 

In Fig.~\ref{fig_1loop} the 1-loop dilepton rate for zero momentum, 
containing such processes, is displayed as a function of the scaled invariant 
mass with the thermal quark mass and is also compared with the Born-rate. 
In the left panel (Fig.~\ref{fig_1loop}($a$)) the van Hove singularities 
due to the nontrivial dispersion 
of quarks in a medium are evident in pole-pole contributions whereas the
pole-cut and cut-cut contributions\footnote{These are due to the space-like
($k^2>k_0^2$) part of the $N$-point HTL functions that acquire a cut 
contribution from below the light cone.} are smooth representing absorption 
and emission of gluons in the medium. The right panel 
(Fig.~\ref{fig_1loop}($b$)) displays the total one-loop contribution 
for a set of values of $g$, where the energy gaps are smoothened due to
the pole-cut and cut-cut contributions. Also the structures due to
the van Hove singularities become also less prominent in the total 
contributions. The HTL rate, in particular, due to the cut contributions 
is also singular at $M\rightarrow 0$ because the
HTL quark-photon vertex is inversely proportional to photon energy.

However, these corrections are not sufficient and two-loop diagrams within 
HTL perturbation scheme contribute to the same order and are even larger 
than the one-loop results~\cite{agkz98}. The total one-
and two-loop rate at ${\vec{\mathbf p}}=0$ and $M<<T$ in the leading 
logarithm, {\em i.e.}, $\ln(1/g)$ approximation reads~\cite{agkz98,agkz99}
\begin{equation}
\frac{dR}{d^4xd^4P}=\frac{5\alpha^2}{9\pi^6}\frac{m_q^2}{M^2} \left [
\frac{\pi^2m_q^2}{4M^2}\ln \frac{T^2}{m_q^2}+ \frac{3m_q^2}{M^2}
\ln\frac{T^2}{m_g^2}+\frac{\pi^2}{4}\ln \left (\frac{MT}{M^2+m_q^2} \right ) 
+2\ln \left (\frac{MT}{M^2+m_g^2} \right ) 
\right ] \ , \label{rate_2loop}
\end{equation}
where the thermal gluon mass is given by $m_g^2=8m_q^2/3$ with 
$m_q=gT/\sqrt 6$. Note that this expression is of the same order in $g$ as 
the Born-term for soft $M\sim gT$. 
Now the Born-term for ${\vec{\mathbf p}}=0$ and $M<<T$ is simply 
given by 
\begin{equation}
\frac{dR}{d^4xd^4P}=\frac{5\alpha^2}{144\pi^4}=1.90\times 10^{-8} \ .
\label{born_p0}
\end{equation}

\begin{figure}[!tbh]
\vspace*{0.44in}
\includegraphics[height=0.47\textwidth, width=0.7\textwidth]{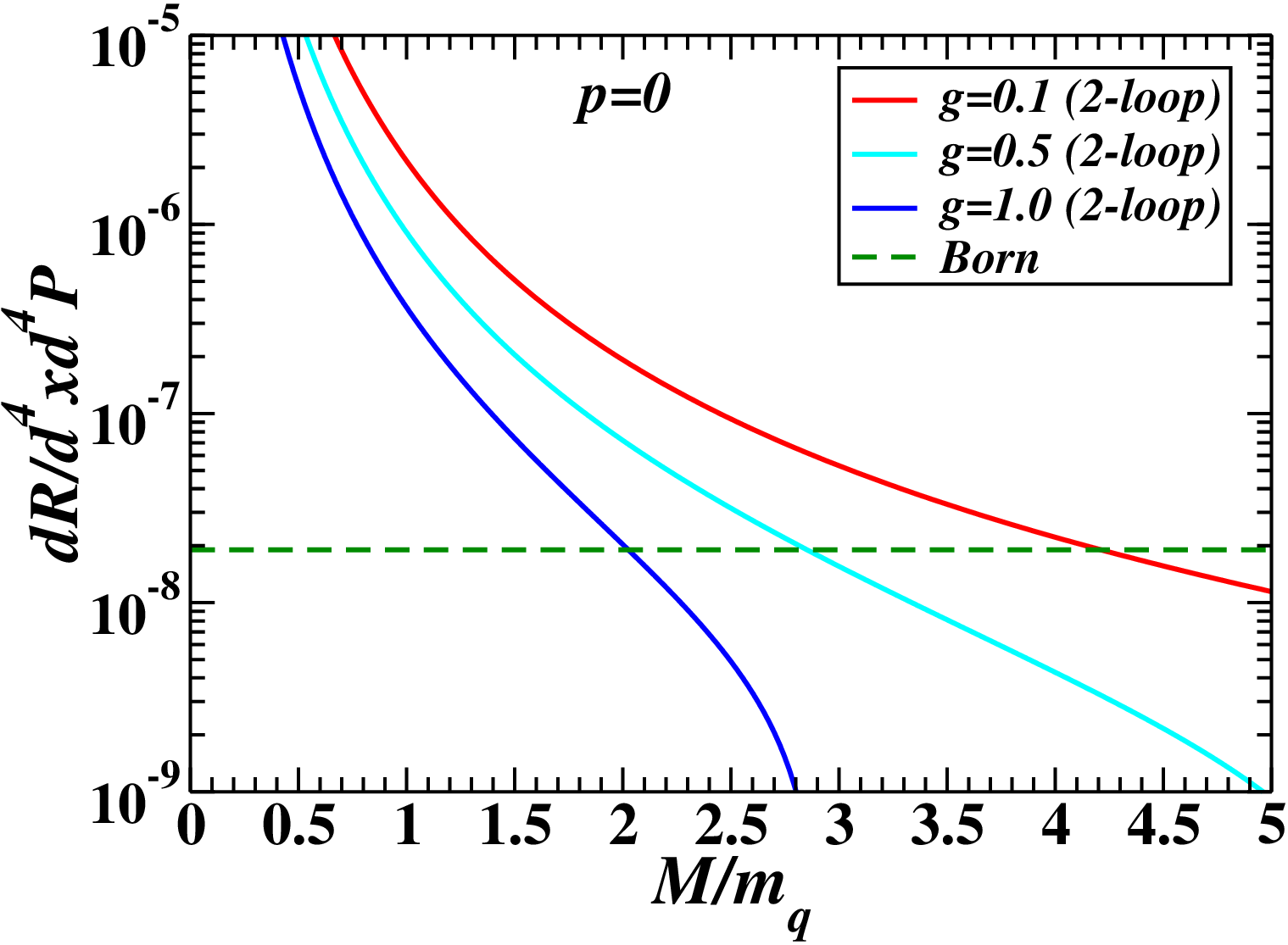}
\vspace*{-0.08in}
\caption{(Color online) Complete 2-loop dilepton rate for small invariant masses $M\sim gT$ 
at zero momentum and Born-rate (dashed line) versus the scaled invariant 
photon mass $M/m_q$ with the thermal quark mass $m_q$.
}
\label{fig_2loop}
\end{figure}

In Fig.~\ref{fig_2loop} the Born-rate and the complete two-loop rate 
for a set of values of $g$ are compared. It is evident from 
Fig.~\ref{fig_2loop} that the $2$-loop rate dominates in the perturbative
regime ($g\leq 1$) over the Born-term for low mass domain, $M/m_q\leq 2$.  
However, the 
van Hove singularities contained in one-loop do not appear as they are 
washed out due to the leading logarithm approximation within the two-loop
HTLpt.

\subsubsection{Semi-hard Rate ($M\sim T$ and $p>> T$)}

For $M$ of the order of $T$ and hard momenta ($p>>T$), the 
$\alpha_s$-correction to the Born-rate has been 
calculated~\cite{tt97} 
within the HTLpt method as
\begin{equation}
\frac{dR}{d^4xd^4P}=\frac{5\alpha^2\alpha_s}{27\pi^3}\frac{T^2}{M^2} e^{-E/T}
\left (\ln \frac{T(m_q+k^*)}{m_q^2}+C\right ) \ ,
\label{rate_alphas}
\end{equation}
where $k^*\approx |Em_q^2/M^2-m_q^2/(4E)| < (E+p)/2$ and $C\approx -0.5$
depends weakly on $M$. In Ref.\cite{cga08} it has been shown that further corrections
to the rate (\ref{rate_alphas}) are necessary. However, numerical results showed only 
a slight modification.

\begin{figure}[!tbh]
\vspace*{0.45in}
\includegraphics[height=0.53\textwidth, width=0.7\textwidth]{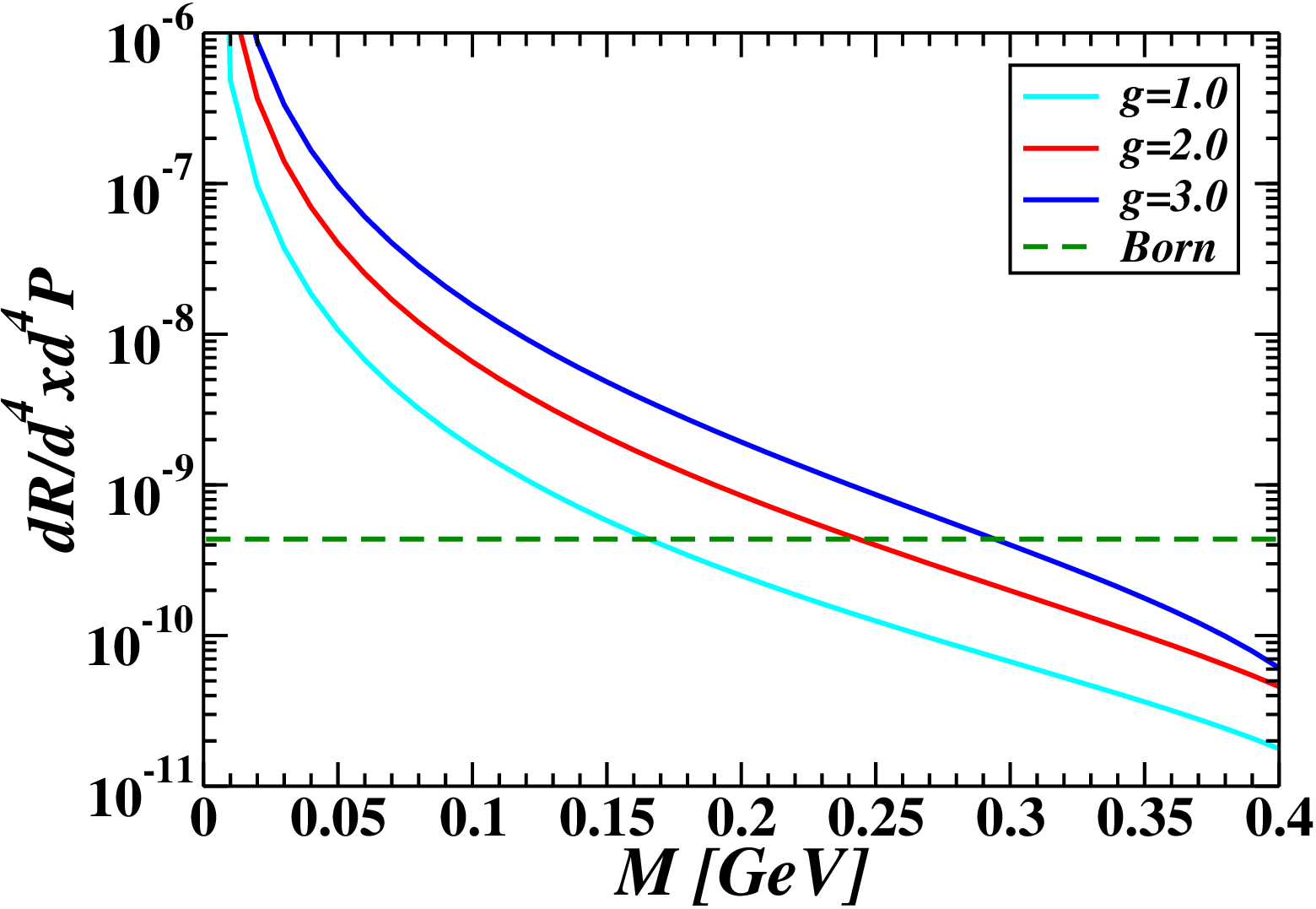}
\vspace*{-0.1in}
\caption{(Color online) $\alpha_s$-correction to the dilepton rate and Born-rate 
(dashed line) versus the invariant photon mass $M$ scaled with the 
thermal quark mass for $T=200$MeV and $E=1$ GeV.}
\label{fig_alphas}
\end{figure}

Assuming  typical values of the strong coupling constant
and temperature, $T=200$ MeV, these corrections dominate over the Born term
for invariant masses below $300$ MeV as shown in Fig.~\ref{fig_alphas}.
Similar results have been obtained using bare quark propagators~\cite{ar92}.
However, the calculation within naive perturbation theory~\cite{cdj94} 
resulted in $\alpha_s$-corrections which are of similar size as the 
Born-rate in the regime $M$ and $p$ of the order of $T$.

\subsubsection{Hard Rate ($M >> T$)}
For $M>> T$ naive perturbation theory using bare propagators and vertices is 
sufficient. This is in contrast to the production of real photons, where one 
encounters an infrared singularity from bare quark propagator~\cite{kls92}. 
For finite $M$, however, this singularity cancels~\cite{aa89}. Bare two-loop 
calculations~\cite{aa89,cdj94} showed that the $\alpha_s$-corrections are 
negligible in this regime. However, a recent calculation of the 
$\alpha_s$-corrections~\cite{kw00} for large invariant mass $M>>T$ and
small momenta $p<<T$ yielded important corrections to the Born-rate for 
invariant masses below $(2-3)T$. However, this work has also been criticized 
\cite{abb00}.

The main problem in applying perturbative results discussed above to realistic
situations is the fact that $g$ is not small but rather we have 
$g\sim 1.5-2.5$. Close to the critical temperature, $T_c$, even $g$ 
could be as high as $6$~\cite{pkps94}. Hence the different momentum 
scales are not distinctly
separated in the real sense and, even if one still believes in perturbative 
results (see Figs.~\ref{fig_1loop} , \ref{fig_2loop} and \ref{fig_alphas}) 
at least qualitatively, it is not clear which of the above rates applies to 
heavy-ion collisions. However, in all cases there are substantial corrections 
to the Born-rate.  The perturbative rates within their uncertainties in 
various regime probably suggest that the Born-rate may not be sufficient 
for describing the low mass dilepton spectrum.  

\subsection{Nonperturbative Rate }

Considering the uncertainty of thermal perturbation theory for QCD a 
nonperturbative approach to the dilepton rate would be desirable. In this
subsection we describe non-perturbative dilepton production rates in a 
deconfined phase in phenomenological models and in a first principle 
calculation, {\em viz.}, within the lattice gauge theory. 

\subsubsection{Rate using Gluon Condensate within the Green Function}

An important issue towards the understanding the phase structure of QCD
is to understand the various condensates, which serve as order parameters
of the broken symmetry phase. These condensates are non-perturbative in nature
and lattice provides a connection with bulk properties of QCD matter. However,
the quark condensate has a rather small impact on the bulk properties,
{\em e.g.,} on the equation of state of QCD matter, compared to the gluon 
condensate~\cite{boyd}. The relation of the gluon condensate to the 
bulk properties such as equation of states, in principle, can be tested 
through hydrodynamic or transport properties sensitive to the equation
of states, but is a non-trivial task.

\begin{figure}[!tbh]
\vspace*{0.47in}
\includegraphics[height=0.6\textwidth, width=0.7\textwidth]{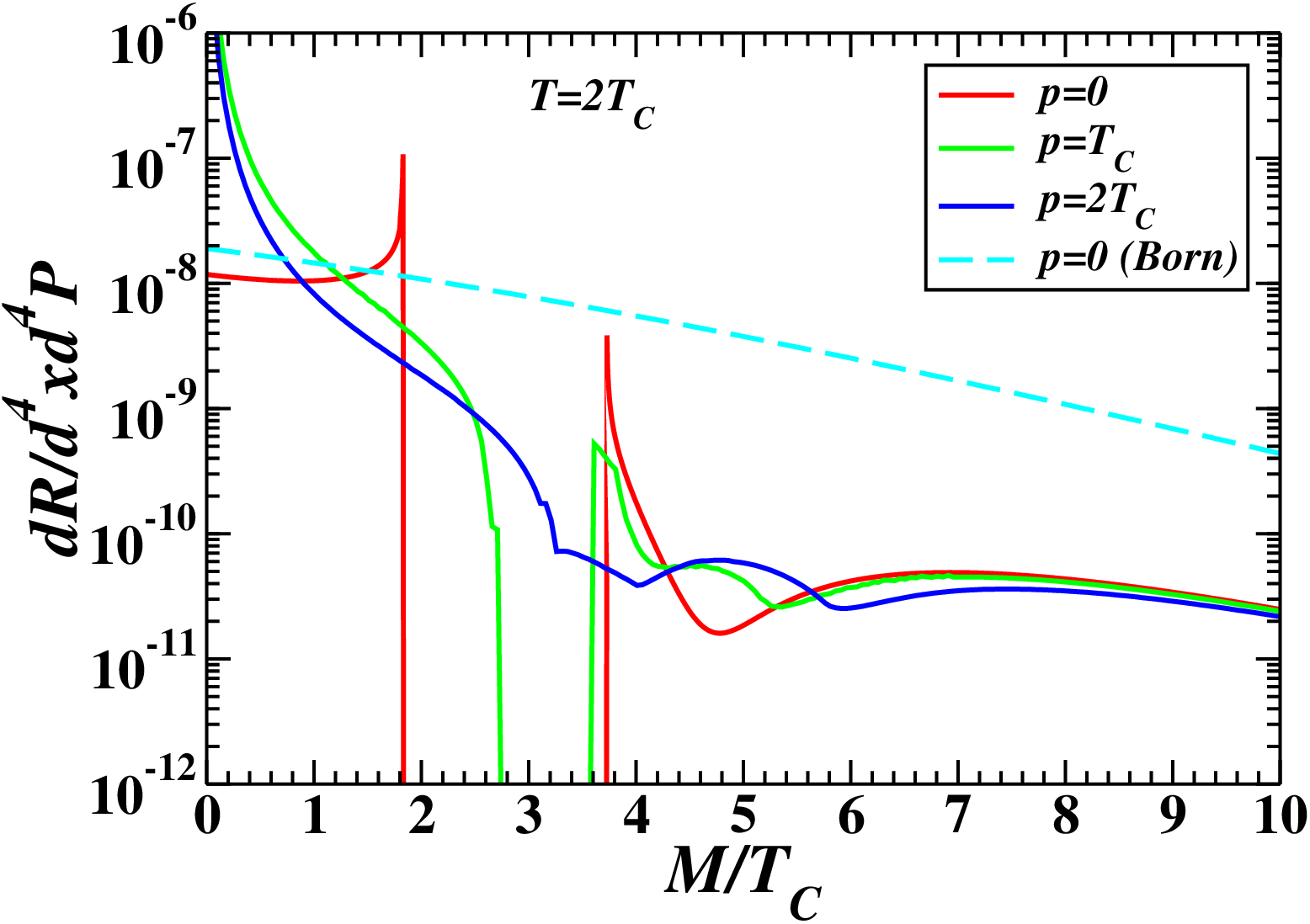}
\vspace*{-0.1in}
\caption{(Color online) Van Hove singularities in the dilepton rate in the presence of gluon
condensate as a function of invariant mass scaled with $T_c$ for a set of 
momenta at $T=2T_c$. The dashed curve is for Born-rate at zero momentum.}
\label{fig_gc}
\end{figure}

A semi-empirical way to consider nonperturbative aspects,{\em e.g}, gluon 
condensate has been suggested by combining lattice results with Green 
function in momentum space~\cite{st99,mst00}. In this approach the effective 
$N$-point functions~\cite{st99,mst00} have been constructed which
contain the gluon condensate in the deconfined phase, 
measured in lattice QCD~\cite{boyd}. The resulting quark dispersion relation 
with a mass $m_q\sim 1.15T_c$~\cite{st99} 
in the medium shows qualitatively the same behaviour as the HTL dispersion, 
leading again to sharp structures (van Hove singularities, energy gap) in 
the dilepton production rates~\cite{mst99}, indicating that this features are
universal in relativistic plasmas independent of the approximation used \cite{markus}.
In Fig.~\ref{fig_gc} the dilepton production rate using gluon condensate
is displayed for various values of momentum at $T=2T_c$ and also compared
with the Born-rate. At very low invariant mass 
($M/T_c\le  2$; for $T_c\sim 165$ MeV, $M\le 330$ MeV) 
with realistic momentum the dilepton rate with gluon condensate dominates 
over the Born-rate. This rate will be important at very low invariant mass as
it has non-perturbative input from lattice QCD that describes the bulk 
properties of the deconfined phase, and is of course free from  
any uncertainty related to the strong coupling $g$ associated with the
perturbative rates discussed in subsec. B.

We, however, also note that the rate deviates from the Born-rate at high
$M/T_c$ ($\ge 4$). The difference at high $M/T_c$ has the origin  
in the asymptotic limit (large momentum $k$) of the quark dispersion
relation with gluon condensates. In this limit it is found that the normal 
quark mode behaves like $w_+=k+c$, where $c$ contains still the non-zero 
contribution from the condensates. The reason for which 
is the use of the momentum independent condensate values. This fact has 
crept in the dilepton rate at high $M/T_c$.  One way out could be to use an 
{\em ad hoc} separation scale ($M/T_c\sim 2-3$) up to which one may employ the 
non-perturbative quark dispersion associated with the gluon condensate
and beyond which a free dispersion is adopted. Alternatively, one could use a momentum 
dependent condensate, which is again beyond the scope of our calculation 
and has to be provided by the lattice analysis. To date we are not aware 
of such analysis. Nonetheless, we note that the nonperturbative contribution 
is important only at low invariant mass as we would see later in sec. III.

\subsubsection{Quark and  $\rho^0$-meson Interaction ($\rho$-meson in QGP)}

We assume that $\rho$-meson like states ($q\bar q$ correlator in the 
$\rho$-meson channel) can exist in a deconfined phase like QGP. Then 
there will also be a contribution from $\rho$-meson channel to the
dilepton pairs ($l^+l^-$) in addition to the perturbative production. 
In order to consider such a channel 
phenomenologically an interaction of $\rho-q$ coupling is 
introduced through the Lagrangian~\cite{tlm00} 
\begin{equation}
{\cal L}= -\frac{1}{4} \rho_{\mu\nu}^a\rho^{\mu\nu}_a
+ \frac{1}{2}m_\rho^2\rho_\mu^a\rho^\mu_a 
+ \bar q \left (i\gamma_\mu\del^\mu -m_q+G_\rho \gamma^\mu \frac{\tau_a}{2}
\rho_\mu^a\right )q ,  \label{lag_rq}
\end{equation}
where $q$ is the quark field, $m_q$ is the quark mass,  $a$ is the isospin
or flavour index, and $\tau_a$ is the corresponding isospin matrix. 
The $\rho - q$ coupling, $G_\rho$, can be obtained in the same spirit as the
$4$-point 
interaction, $G_2(\bar q\gamma_\mu\tau_a q)^2$, in NJL-model. This suggests
$G_\rho=\sqrt{8m_\rho^2G_2} \sim 6 $, by taking $G_2$ from the literature. The
similar value for $G_\rho$ can be obtained by simply assuming that the $\rho$-meson
couples in a universal way to nucleons, pions and quarks~\cite{tlm00}.

Now using the Vector Meson Dominance (VMD)~\cite{gale} the photon self-energy 
is related to the $\rho^0$ meson propagator, $D_{\mu\nu}(P)$, by
\begin{equation}
{\mbox{Im} \Pi_\mu^\mu(P)}= \frac{e^2}{G_\rho^2} m_\rho^4 \ {\mbox{Im} 
D_\mu^\mu(P)} \ \ . \label{vdm_rel}
\end{equation}

Then the thermal dilepton production rate from the $\rho$-meson 
can be written as
\begin{equation}
\frac{dR}{d^4x \, d^4P}  \ = -\ \frac{1}{3\pi^3} \ \frac{\alpha^2}
{G_\rho^2} \ \frac{m_\rho^4}{M^2} \  \ \frac{1}{e^{E_p/T}-1} \ 
\left ({\cal A}_\rho^L + 2 {\cal A}_\rho^T \right) \ \ , \label{rate_rpp}
\end{equation}
and the spectral functions for $\rho$-meson can be obtained from the
self-energy of $\rho-$meson as
\begin{eqnarray}
{\cal A}_\rho^L(P) &=& \frac {{\rm {Im}} {\cal F}} { \left ( M^2 -m_\rho^2 - 
{\rm {Re} {\cal F}}\right )^2 + ({{\rm {Im}} {\cal F}})^2}  \ ,\\
{\cal A}_\rho^T (P) &=& \frac {{\rm {Im}} {\cal G}} { \left ( M^2 -m_\rho^2 - 
{\rm {Re} {\cal G}}\right )^2 + ({{\rm {Im}} {\cal G}})^2}\ ,
\end{eqnarray}
where ${\cal F}= - \frac{P^2}{p^2}\Pi^{00}(P)$ and ${\cal G}=\Pi_T(P)$ with
$L$ and $T$ stand for longitudinal and transverse modes, respectively.

Going beyond the HTL approximation, the integral
expression for the matter part of the one-loop photon self energy 
for assymetric charges in the deconfined phase ({\em viz.}, with  non-zero 
chemical potential, $\mu$, which would be appropriate for FAIR 
energies~\cite{fair}) can be obtained easily by extending the results 
of Ref.~\cite{tlm00} to finite $\mu$ as,
\begin{eqnarray}
{\mbox{Re}\ {\cal F}} &=& 
\frac{3G^2}{4\pi^2}\frac{M^2}{p^2} \int \limits_{0}^\infty dk \ k
\left[n(\omega_k-\mu)+n(\omega_k+\mu) \right]\left (
-2 \frac{k}{\omega_k} + \frac{M^2+4\omega_k^2}{4p\omega_k}
 \ln |a| + \frac{p_0}{p} \ln |b| 
\right ) \ , \nonumber \\
{\mbox{Im}\ {\cal F}} &=&
\frac{3G^2}{4\pi}\frac{M^2}{p^3} \int \limits_{k_-}^{k_+} dk \ k
\left[n(\omega_k-\mu)+n(\omega_k+\mu) \right]\left ( p_0-\omega_k
-\frac{M^2}{4\omega_k}\right )
\ , \nonumber \\
{\mbox{Re}\ {\cal G}} &=&
\frac{3G^2}{4\pi^2} \int \limits_{0}^\infty dk \ \frac{k^2}{\omega_k} 
\left[n(\omega_k-\mu)+n(\omega_k+\mu) \right]\left (
-\left[ \frac{\omega_k^2M^2}{2p^3k}+\frac{M^2}{4pk}
+\frac{M^4}{8p^3k}+\frac{m_q^2}{2pk} \right]
\ln |a| \right.\nonumber \\
&&\left. \ \ \ \ \ \ \ \ \ \ \ \ \ \ \ \ \ \ \ \ \ \  \ \ \ \ \ \ \ 
\ \ \ \ \ \ \ \ \ \ \ \ \ \ \ \ \  \ \ \ \ \ \ \ 
-\frac{p_0M^2\omega_k}{2p^3k} \ln |b| 
 +\frac{M^2}{p^2} +2 \right ) \nonumber \\
{\mbox{Im}\ {\cal G}} &=& 
\frac{3G^2}{8\pi p} \int \limits_{k_-}^{k_+}  dk k
\ \ \left[n(\omega_k-\mu)+n(\omega_k+\mu) \right] \ \
 \left (
-\omega_k+\frac{m_q^2}{\omega_k}+\frac{p_0^2}{p^2}\omega_k 
 +\frac{M^2}{2\omega_k}
\right. \nonumber \\ 
&&\left. \ \ \ \ \ \ \ \ \ \ \ \ \ \ \ \ \ \ \ \ \ \ \ \ \ \ \ \ \
 \ \ \ \ \ \ \ \ \ \ \ \ \ \ \ \ \ \ \ \ \ \ \ \ \ \ \ \ \
+\frac{M^4}{4\omega_kp^2}-\frac{p_0M^2}{p^2}
\right ) \ , \label{rate_rq} 
\end{eqnarray}
along with 
\begin{eqnarray}
a \ = \ \frac{(M^2+2pk)^2 - 4p_0^2\omega_k^2}
{(M^2-2pk)^2 - 4p_0^2\omega_k^2} \ \ , &&
b \ = \ \frac{M^4 - 4(pk+p_0\omega_k)^2}
{M^4 - 4(pk-p_0\omega_k)^2} \ \ , \nonumber \\
k_- \ = \frac{1}{2}\left | p_0 \sqrt{1-\frac{4m_q^2}{M^2}} 
- p \right| \ , &&
k_+ \ =  \frac{1}{2}\left ( p_0 \sqrt{1-\frac{4m_q^2}{M^2}} + p\right ) \ ,
\nonumber
\end{eqnarray}
where $\omega_k=\sqrt{k^2+m_q^2}$.

\begin{figure}[!tbh]
%\vspace*{0.32in}
\subfigure{
\includegraphics[height=0.44\textwidth, width=0.45\textwidth]{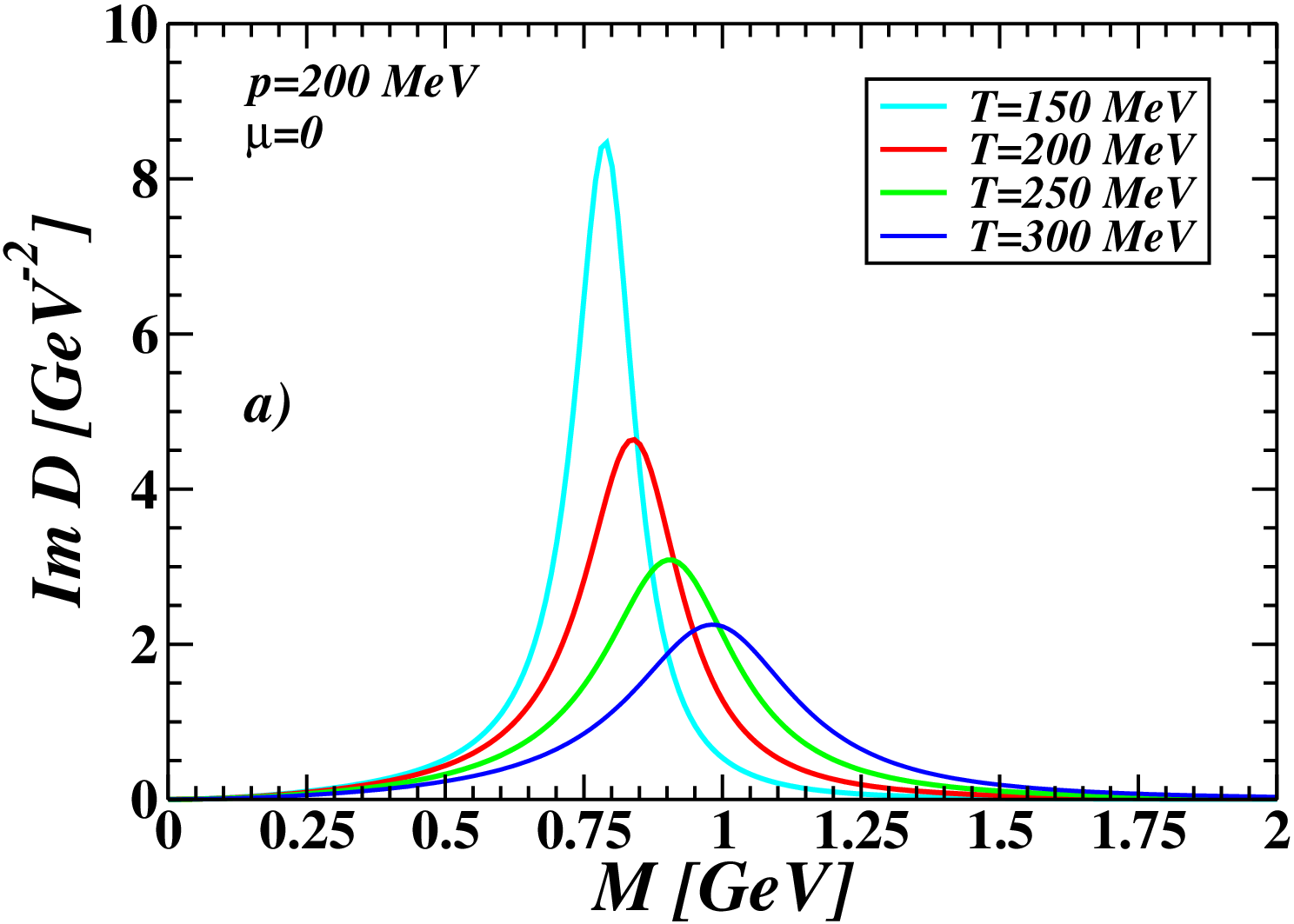}}
\subfigure{
\includegraphics[height=0.44\textwidth, width=0.45\textwidth]{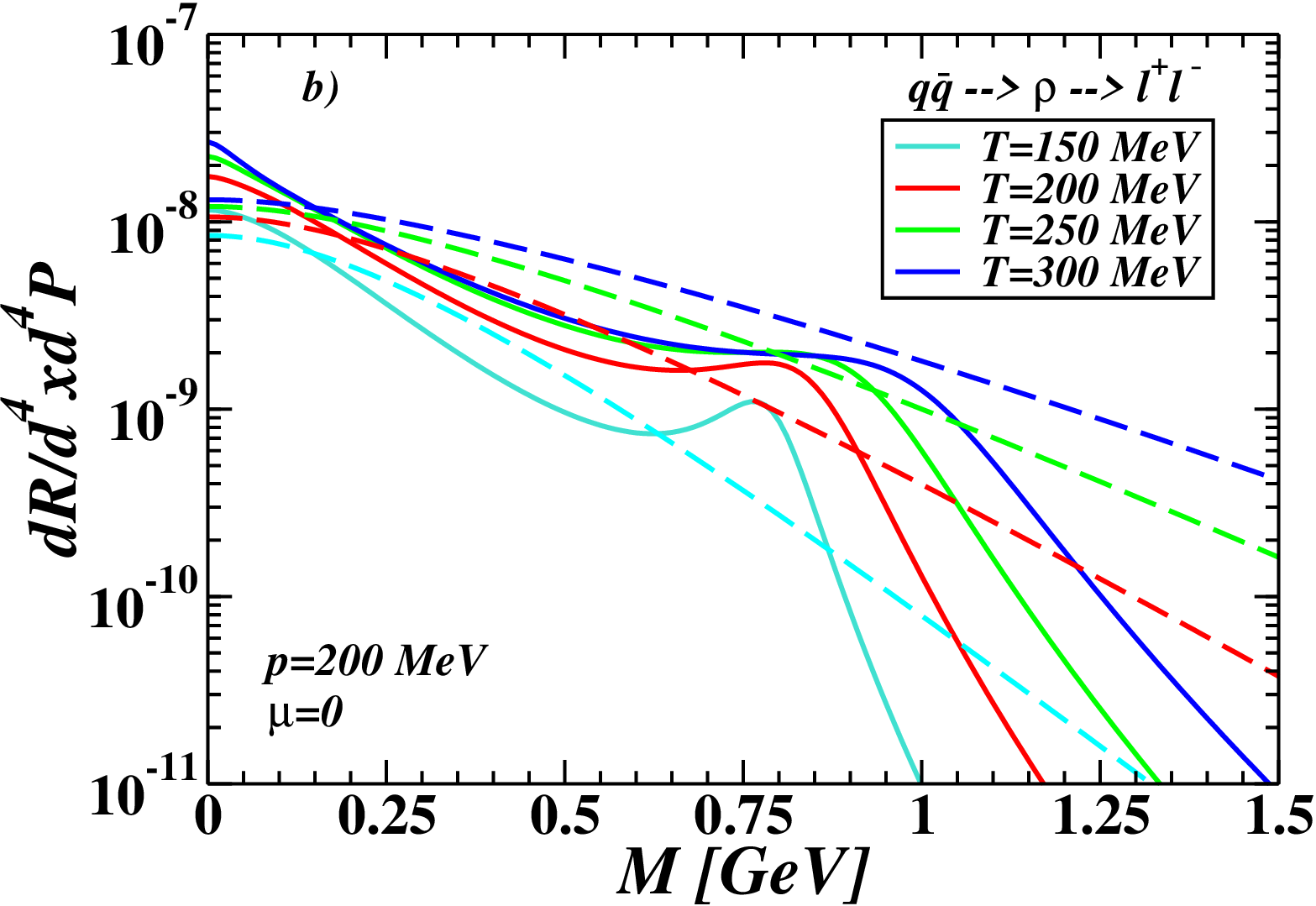}}
\vspace*{-0.12in}
\caption{(Color online){\em Left panel:} Imaginary part of $\rho$-meson propagator
(spectral function) as a function of the invariant mass $M$ for a set 
of values $T$. {\em Right panel:}
The dilepton rate from $\rho$-meson in a QGP as a function of 
$M$. The dashed lines are corresponding Born-rates. We have used 
$G_\rho=6$.} 
\label{fig_rho_t}
\end{figure}

In Fig.~\ref{fig_rho_t} the $\rho$-meson spectral function related to
the imaginary part of the $\rho$-meson propagator (left panel) in 
(\ref{vdm_rel}) and the dilepton rate (right panel) are displayed for 
various temperature with $\mu=0$ and $p=200$ MeV. As the temperature 
increases the peak in the imaginary part of the $\rho$-meson propagator 
$D$ becomes broader and is also reflected in the dilepton rate. 
In the low mass region ($\leq 1$ GeV) the rate is comparable with the 
Born-rate.

\begin{figure}[!tbh]
\vspace*{0.30in}
\subfigure{
\includegraphics[height=0.44\textwidth, width=0.44\textwidth]{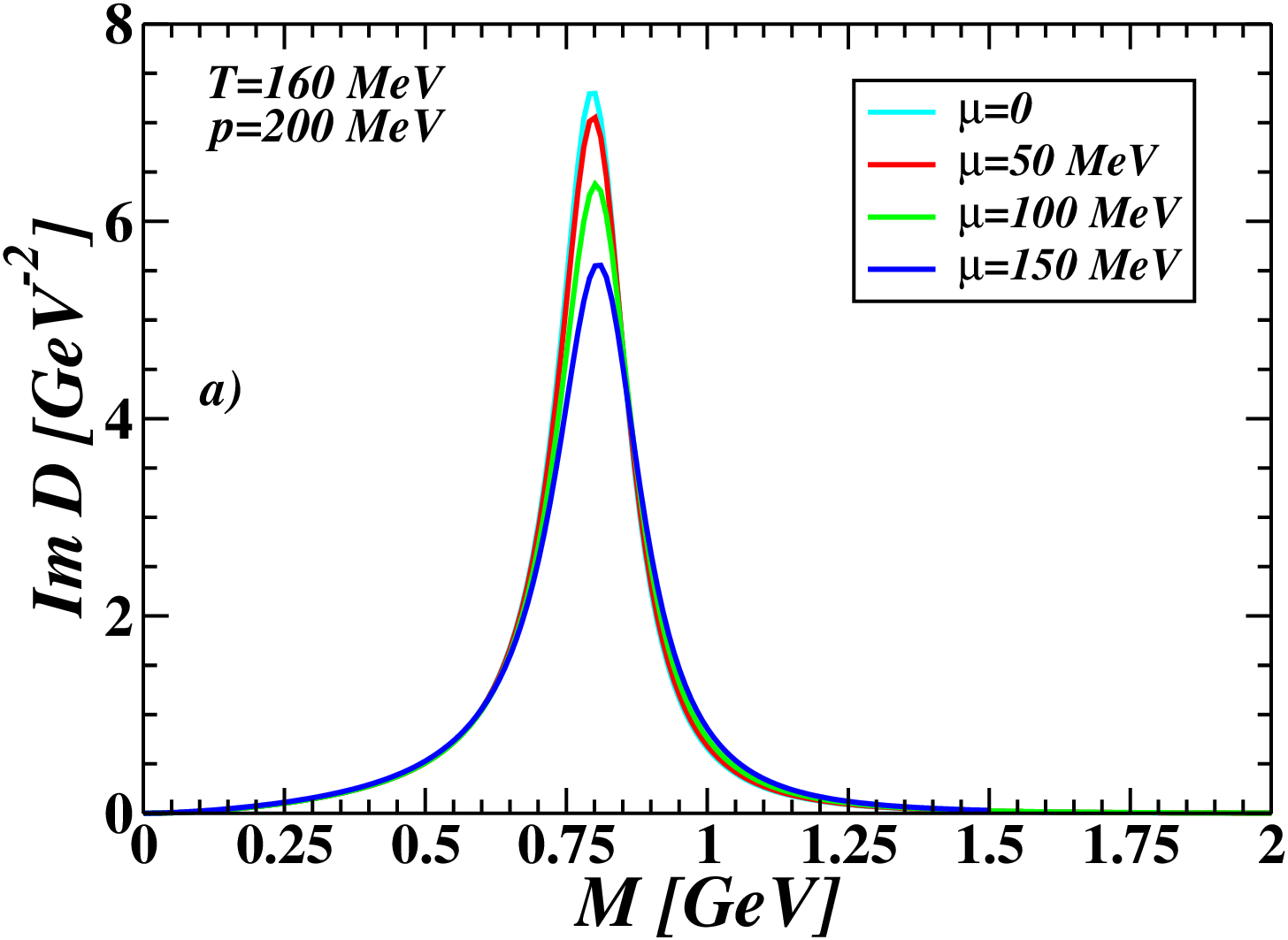}}
\subfigure{
\includegraphics[height=0.44\textwidth, width=0.45\textwidth]{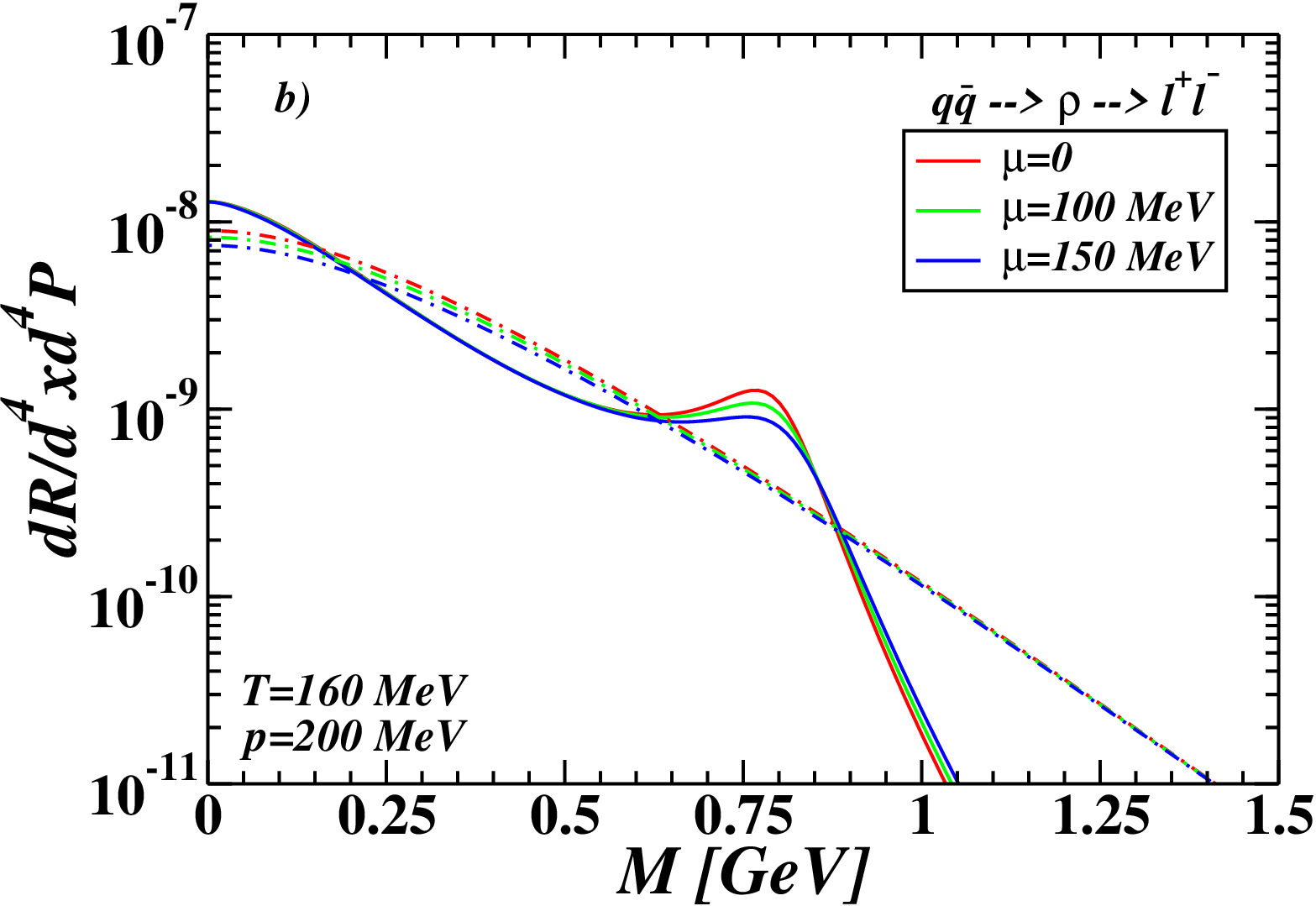}}
\vspace*{-0.13in}
\caption{(Color online) Same as Fig.~\ref{fig_rho_t} but for different $\mu$ at 
a given $T$.}
%\caption{{\em Left panel:} Imaginary part of $\rho$-meson propagator
%(spectral function) as a function of $M$ for a set of values of $\mu$. 
%{\em Right panel:}
%The dilepton rate from $\rho$-meson in a QGP as a function of 
%$M$. We chose $G_\rho=6$. The dash-dotted lines are corresponding Born-rates.} 
\label{fig_rho_mu}
\end{figure}

In Fig.~\ref{fig_rho_mu} the $\rho$-meson spectral function (left panel)
and the dilepton rate (right panel) are displayed for various $\mu$ at
$T=160$ MeV and $p=200$ MeV, which could be appropriate in the perspective 
of FAIR energies. The effect of broadening of the $\rho$-meson is far less
pronounced with increasing $\mu$ than increasing $T$, indicating that the
$\rho$-meson is not completely melted in the case of a system with finite
baryon density such as expected at FAIR energies even above the phase 
transition.  However, dilepton rates from 
$\rho$-meson as shown in Figs.~\ref{fig_rho_t} and \ref{fig_rho_mu} are 
comparable with the Born-rate in QGP in the low mass region ($M\le 1$ GeV), 
may be an indication for chiral restoration~\cite{rapp1,rapp2,tlm00}. 
In addition this rate would be
important for invariant masses below $1$ GeV.

We also note that if one includes higher mass vector mesons such as 
$\phi$-meson within VMD, then there will be a peak corresponding to an
invariant mass of the order of $\phi$-meson mass but in low mass region
($M\leq 1$ GeV) there should be a very little change (less than $5\%$) in the
dilepton rate. Since we are interested in the low mass region, we have 
not discussed $\phi$-meson here.
\subsubsection{Rate from Lattice Gauge Theory}

The thermal dilepton rate describing the production of lepton pairs
with energy $\omega$ and momentum ${\vec {\mathbf p}}$ is related to the
Euclidian correlation function~\cite{munshi} of the vector current, 
$J_V^\mu={\bar {\psi}}(\tau,{\vec {\mathbf x}})\gamma^\mu \psi(\tau,{\vec 
{\mathbf x}})$, which can be calculated numerically in the framework of lattice
gauge theory. The thermal two-point vector correlation function in coordinate
space, ${\cal G}_V(\tau, {\vec {\mathbf x}})$, is defined as
\begin{equation}
{\cal G}_V(\tau, {\vec {\mathbf x}})=\langle J_V(\tau, {\vec {\mathbf x}})
J_V^\dagger(\tau, {\vec {\mathbf x}})\rangle =T\sum_{n=-\infty}^{\infty} \int
\frac{d^3p}{(2\pi)^3} e^{-i(w_n\tau-{\vec{\mathbf p}}
\cdot {\vec{\mathbf x}})} \chi_V(w_n,{\vec{\mathbf p}}) \ , \label{vec_cor}
\end{equation}
where the Euclidian time $\tau$ is restricted to the interval 
$[0,\beta=1/T]$, and 
the Fourier transformed correlation function $\chi_V$ is given at the discrete 
Matsubara modes, $w_n=2\pi n T$. The imaginary part of the momentum space
correlator gives the spectral function $\sigma_V(\omega, {\vec{\mathbf p}})$, as
\begin{equation}
\chi_V(w_n,{\vec{\mathbf p}}) = - \int_{-\infty}^{\infty} 
\frac{\sigma_V(\omega,{\vec{\mathbf p}})}{iw_n-\omega+i\epsilon} \ \Rightarrow \
\sigma_V(\omega,{\vec{\mathbf p}})= \frac{1}{\pi} {\mbox{Im}}
\ \chi_V(\omega,{\vec{\mathbf p}}) \ . \label{spec_mom}
\end{equation}
Using (\ref{vec_cor}) and (\ref{spec_mom}) the spectral representation of
the thermal correlation functions at fixed momentum in coordinate space 
can be obtained as
\begin{equation}
{\cal G}(\tau, {\vec {\mathbf p}})=\int_0^\infty\ d\omega \ 
\sigma_V (\omega, {\vec {\mathbf p}}) \
\frac{\cosh[\omega(\tau-\beta/2)]}{\sinh[\omega\beta/2]} \ . \label{corr_mom}
\end{equation}

The vector spectral function, $\sigma_V$, is related to the differential
dilepton production rate~\cite{munshi}\footnote{A factor of $2$ differs 
from that of Ref.~\cite{karsch}} as
\begin{equation}
\sigma_V(\omega, {\vec {\mathbf p}})=\frac{18\pi^2N_c}{5\alpha^2} \omega^2 
\ (e^{\omega/T}-1)\ \frac{dR}{d^4xd^4P} \ , \label{rel_dilep_spec}
\end{equation}
where $N_c$ is the number of color degree of freedom.

\begin{figure}[!tbh]
\vspace*{0.45in}
\includegraphics[height=0.6\textwidth, width=0.7\textwidth]{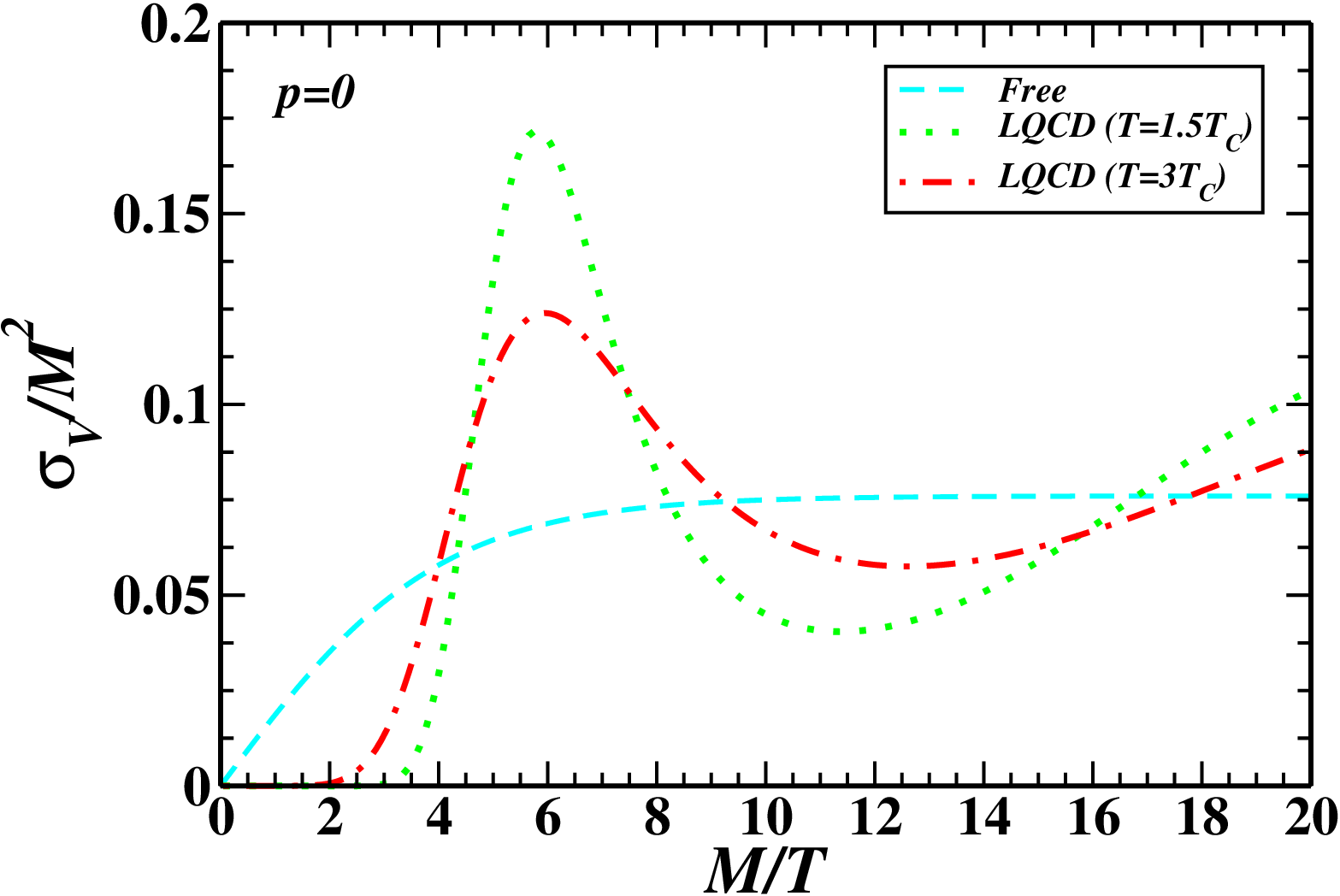}
\vspace*{-0.1in}
\caption{(Color online) The zero momentum (${\vec {\mathbf p}}=0$) vector spectral 
function, reconstructed from the correlation function~\cite{karsch} within 
lattice gauge theory in quenched QCD using MEM, scaled with $M^2$ as 
a function of $M/T$ compared with that of the free one above the deconfinement
temperature $T_c$.}
\label{fig_spec_lat}
\end{figure}

A finite temperature lattice gauge theory calculation is performed on
lattices with finite temporal extent $N_\tau$, which provides information
on the temporal correlation function, ${\cal G}(\tau, {\vec {\mathbf p}})$,
only for a discrete and finite set
of Euclidian times $\tau =k/(N_\tau T), \ \ k=1,\cdots \ N_\tau$.
The correlation function, ${\cal G}(\tau, {\vec {\mathbf p}})$, has been
computed~\cite{karsch} within the quenched approximation of QCD using 
non-perturbative improved clover fermions~\cite{clov} through a probabilistic 
application based on the maximum entropy method (MEM)~\cite{mem} for 
temporal extent $N_\tau= 16$ and 
spatial extent $N_\sigma = 64$. Then by inverting the integral in 
(\ref{corr_mom}), the spectral function is reconstructed~\cite{karsch}
 in lattice QCD. In Fig.\ref{fig_spec_lat} such a reconstructed spectral 
function scaled with $M^2$ (equivalently $\omega^2$ for $\vec{\mathbf p}=0$) 
is displayed 
as a function of $M/T$. The vector spectral functions above the deconfinement 
temperature ({\em viz.}, $T=1.5T_c \ {\mbox{and}} \ 3T_c$) show an 
oscillatory behaviour compared to the free one. The spectral functions are
also found to be vanishingly small for $M/T\le 4$ due to the sharp cut-off 
used in the reconstruction. 

\begin{figure}[!tbh]
\vspace*{0.49in}
\includegraphics[height=0.6\textwidth, width=0.7\textwidth]{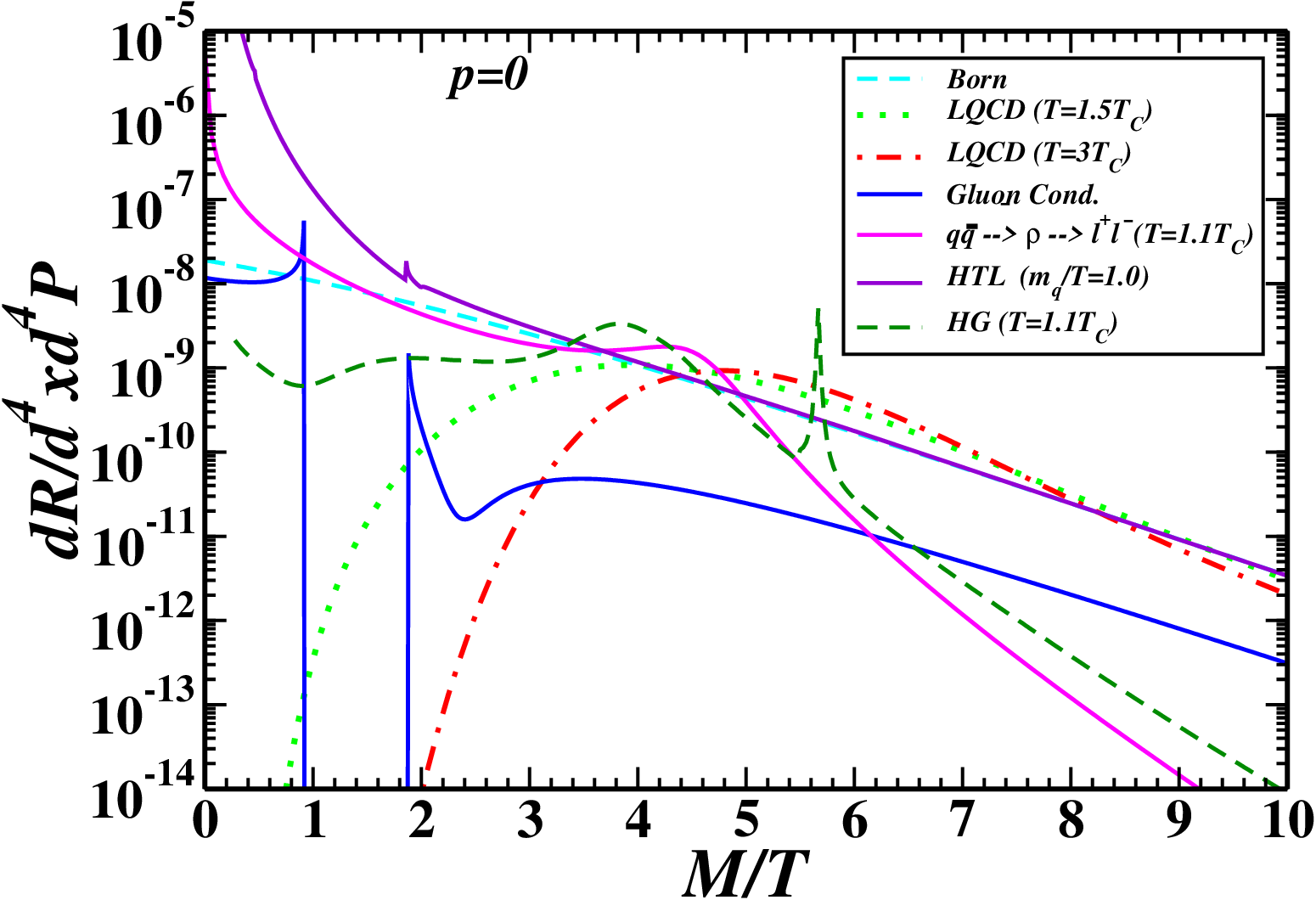}
\vspace*{-0.1in}
\caption{(Color online) Comparison of various dilepton rates in a QGP 
and in a hadron gas (HG) as a function of 
$M/T$ for momentum ${\vec {\mathbf p}}=0$. The  critical temperature
is $165$ MeV~\cite{peter} and the value of $G_\rho$ is chosen as 6. 
The in-medium HG rate is from the recent calculations of Ref.\cite{gsa10}.}
\label{fig_comp}
\end{figure}

A direct calculation of the differential dilepton rate 
using (\ref{rel_dilep_spec}) above the deconfined temperature ($T_c$)
at ${\vec{\mathbf p}}=0$ was first time done in Ref.\cite{karsch} within 
the lattice gauge theory in quenched QCD using the MEM.
In Fig.~\ref{fig_comp} the lattice dilepton rates at ${\vec {\mathbf p}}=0$
for two temperatures ($T=1.5T_c$ and $3T_c$) are displayed as a function of
the scaled invariant mass with temperature and $M/T=\omega/T$, the energy of 
the dileptons. We have also compared the  perturbative, non-perturbative 
and in-medium hadrons rates within the same normalisation as shown in 
the plot. We note that the rate with gluon condensate perfectly scales with
the temperature whereas that of HTL one depends on the choice of the
effective coupling, $m_q/T\sim g/\sqrt 6$. 
The lattice results 
are comparable within a factor of $2$ with the Born-rate as well as that of 
HTLpt at high invariant mass $M/T\ge 4$. The absence of peak structures 
around the $\rho$-mass and also at higher $M$ in the lattice dilepton rate 
probably constrain the broad resonance structures in the dilepton rates. 
However, for invariant mass 
below $M/T\le 4$ the lattice dilepton rate falls off very fast. This is due 
to the fact that the sharp cut-off is used to reconstruct the spectral function 
from the correlation function and the finite volume restriction in the lattice 
analysis. The lattice analysis is also based on rather small statistics.  
These lattice artefacts are related to the smaller invariant masses which 
in turn indicate that it is not yet very clear whether there will be any 
low mass thermal dileptons from the deconfined phase within the lattice 
gauge theory calculation. Future analysis could improve the situation in 
this low mass regime. One cannot rule out~\cite{karsch} the existence of 
van Hove singularities and energy gap, which are general features of 
massless fermions in a relativistic plasma~\cite{markus}, in the low 
mass dileptons. This calls for a further investigations on the lattice
gauge theory side by improving and refining the lattice ingredients and
constraints.

On the other hand, in HTLpt, apart from the 
uncertainty in the choice of $g$, the low mass ($M\rightarrow 0$, 
vanishing photon energy) one-loop dilepton rate obtained from vector
meson spectral function analysis~\cite{munshi} diverges because the
quark-photon 
vertex is inversely proportional to the photon energy. This also requires 
a further improvement of the HTLpt. However, we assume that 
the perturbative rate could also be reliable for $M\ge 200$ MeV 
with $T\ge 200$ MeV and $g\ge 2$. 
The other two phenomenological models, {\em viz.}, gluon condensate 
measured in 
lattice~\cite{boyd} and $\rho-q$ interaction in the deconfined phase 
as discussed respectively above in subsec. C 1 and 2, 
for non-perturbative dilepton production at low mass regime are at 
least cleaner than the perturbative rates which depend weakly on the
choice of the strong coupling constant. The rate with gluon condensate
is free from strong coupling whereas that from $\rho-q$ interaction does not
depend strongly on the choice of the coupling (see below in 
Fig.~\ref{fig_mom_intg}).  
In addition to the perturbative rate 
these two together could also provide a realistic part of the dilepton 
rate at low mass regime ($\le 1$ GeV) from the deconfined phase, as
also can be seen in the next section. As a 
comparison, we have also shown the recent rate from in-medium hadrons 
of Ref.~\cite{gsa10}, where the analytic structure of $\rho$-meson 
propagator has been used due to its interaction with thermal mesons.

\section{Momentum Integrated Rate}
The momentum integrated dilepton rate can be obtained as
\begin{equation}
\frac{dR}{d^4xdM^2} = \int \frac{d^3p}{2p_0} \ \frac{dR}{d^4xd^4P} \ .
\label{mi}
\end{equation}

\begin{figure}[!tbh]
\vspace*{0.49in}
\includegraphics[height=0.6\textwidth, width=0.7\textwidth]
{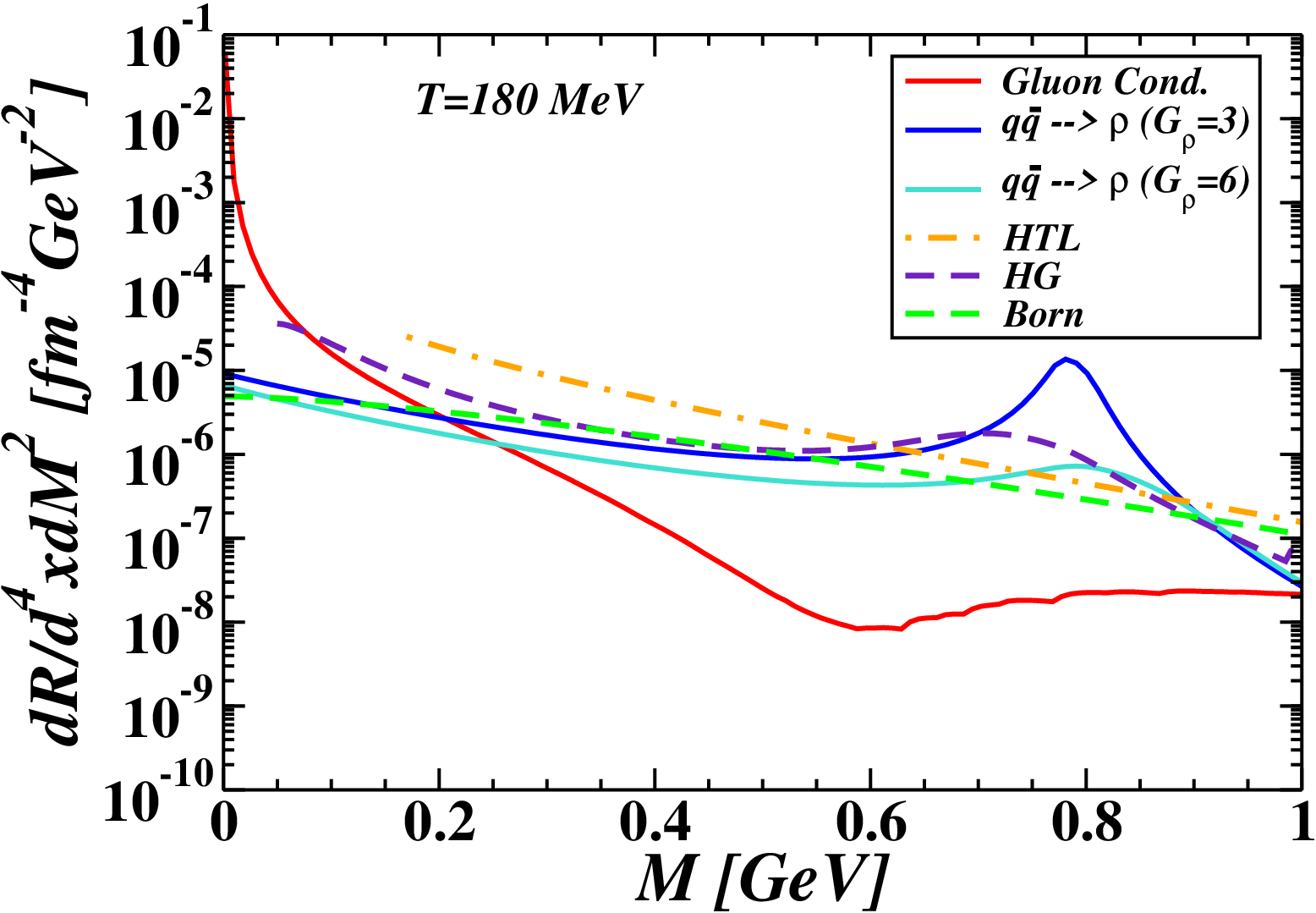}
\vspace*{-0.1in}
\caption{(Color online) Momentum integrated dilepton rate as a function of the invariant mass 
$M$. We have used $T_c=165$ MeV for the nonperturbaive rate with gluon 
condensate. The in-medium hadronic rate (HG) is from Ref.~\cite{gsa10}.}
\label{fig_mom_intg}
\end{figure}

In Fig.~\ref{fig_mom_intg} dilepton rates from QGP and in-medium 
hadrons
are displayed as a function of invariant mass. As can be seen the 
non-perturbative contribution using gluon condensate dominates over 
Born-rate as well as the perturbative rate below $M\le 200$ MeV.
The non-perturbative rate is indeed important with input from
the first principle calculations~\cite{boyd} that describe the bulk
propertirs of the deconfined phase. More importantly, this domain is 
also beyond reach of any reliable perturbative calculations in true
sense. The rate from $\rho-q$ interaction is almost of the same order as 
that of the Born-rate as well as the in-medium hadrons for $M\le 600$ MeV
 whereas it is higher than the perturbative one in the domain 
$600\le M (\mbox{MeV})\le \ 800$  due to the broadening of the $\rho$ peak 
in the medium.  We also note that this rate has a weak dependence on the 
realistic range of values of the $\rho-q$ coupling $(2-6)$.  In addition the 
higher order perturbative rate from HTL, as discussed above, becomes 
reliable for $M\ge 200$ MeV and also becomes of the order of 
Born-rate for $M\ge 500$ MeV. We also note that the momentum integrated 
HTL rate used here has been obtained recently 
by Rapp et al.~\cite{rapp2} through a parametrization of the prefactor of 
the zero momentum 1-loop HTL rate~\cite{yaun} with a temperature dependent $g$, 
which is claimed~\cite{private} to reproduce the Born-rate
 in (\ref{born_gen}) within the appropriate limit.
Now for a comparison, we have also shown the recent 
rate from the in-medium hadrons of Ref.~\cite{gsa10}. 
It is now clear that for low invariant mass ($\le 1$ GeV) only 
the Born-rate from the QGP is not realistic as well as insufficient 
for describing the dilepton rate. 
Instead we suggest that the non-perturbative rate with gluon condensate 
should be important for $M\le 200$ MeV whereas the rates from 
$\rho-q$ interaction and HTLpt are important for $M\ge 200$ MeV.
Below we discuss some aspects
of the quark-hadron duality hypothesis~\cite{qhdual}.

\section{Thoughts on the Quark-Hadron Duality Hypothesis}

It is advocated~\cite{rapp1,qhdual} that due to the potential 
broadening of the $\rho$-meson resonance suffering in a dense hadronic 
environment the overall (momentum integrated) dilepton rate out of 
the hadronic gas becomes equivalent to that from deconfined phase 
as 
\begin{equation}
\frac{dR_H}{d^4xdM^2} \approx  \frac{dR_Q}{d^4xdM^2} \ \ \ , \label{duality_rw}
\end{equation}
which entails a reminiscence to a simple perturbative $q\bar q$ 
annihilation in the vicinity of the expected QGP phase transition.
This hypothesis of 'extended' quark-hadron duality for the thermal 
source of low mass dileptons has been claimed as an indication for 
chiral symmetry restoration~\cite{rapp1,rapp2,qhdual} in the
deconfined phase.  
However, we would like to note that in this hypothesis the volume of 
QGP and hadronic gas was assumed to be same in a given instant of time 
and therefore, the dileptons shine equally bright from  both phases 
at a given instant of time per unit volume. This denotion of quark-hadron 
duality should be carefully re-addressed on its general validity, as the 
suggestive conclusion is indeed far-reaching. A more realistic way to 
look into it is envisaged below.

The momentum integrated rate in (\ref{mi}) shall be gauged to the adequate 
degrees of freedom in a particular phase. A certain measure is given by
the corresponding entropy density. Hence we suggest that for duality to hold
one approximately should have
\begin{equation}
\frac{1}{s_H}\frac{dR_H}{d^4xdM^2} \approx  
\frac{1}{s_Q} \frac{dR_Q}{d^4xdM^2} \ \ \ , \label{duality_us}
\end{equation}
where $s_i\ (i=H,Q)$ is the entropy density of the respective phase.
For an isoentropic crossing over the phase transition, one has 
$s_H dV_H \approx s_Q dV_Q$. Hence if one takes into account the 
respective volume of both phases at a given
instant of time, then instead of (\ref{duality_rw}) one should ask for 
\begin{equation}
{dV_H}\frac{dR_H}{d^4xdM^2} \approx {dV_Q} \frac{dR_H}{d^4xdM^2}
\ \ \ , \ \label{duality_carst}
\end{equation}
where $dV_i$ ($i=Q, \ H$) is the volume of the respective phase. 
Now, at a given instant of time this can lead to
\begin{equation}
%dS_Q \approx dS_H
\frac{dR_H}{dtdM} \approx  \frac{dR_Q}{dtdM}
\ \ , 
\label{duality_carst1}
\end{equation}
where ${dR_i/dtdM}$ is the total yield per time from total phase $i$ in
the system at any instant of time. Therefore, equation (\ref{duality_carst1}) 
means that the fireball emits the same number of dileptons per unit time
either if described by a hadronic or by a deconfined partonic description. 
This could likely be a more realistic way to look into
the quark-hadron duality.
Now, even if the momentum integrated rates in (\ref{mi}) from both
phases are same in some kinematic domain ({\em e.g.,} see 
Fig.~\ref{fig_mom_intg}) may not necessarily imply a quark-hadron duality
as given by (\ref{duality_carst1}) because hadronic volume is expected to 
be larger than that of QGP by at least a factor of 4 to 5. Furthermore, 
we also note that the quark-hadron duality should also be true for any 
momentum at a given instant of time.

\section{Conclusion}

We have discussed the low mass dilepton production rate from the 
deconfined phase within various models, {\em viz.}, perturbative and 
non-perturbative, and compared with that of first principle 
calculations based on lattice gauge theory and in-medium hadrons.
We also have discussed in
details the limitations and uncertainties of all those models at
various domains of the invariant mass. It turns out that at very 
low invariant mass ($\le 200$ MeV) the non-perturbative rate using 
gluon condensate measured in lattice becomes important as this
domain is beyond reach of any reliable perturbative calculations. 
The other non-perturbative contribution from $\rho-q$ interaction
also becomes important below $1$ GeV as it is almost of same order as those 
of the Born and in-medium hadrons. We also note that these two rates
are at least cleaner than the perturbative
rates, in the sense that the gluon condensate rate has non-perturbative 
input from lattice equation of states and is thus free from any coupling 
uncertainties whereas the $\rho-q$ interaction rate does not depend strongly
on the choice of its coupling.  We also 
discussed the $\rho-q$ interaction in the perspective of FAIR scenario. 

On the other hand the perturbative contribution, within its various 
uncertainties, becomes steady and reliable beyond $M> 200$ MeV and 
also becomes comparable with the Born-rate and the LQCD rate for 
$M\ge 500$ MeV. The LQCD rate also constrains the broad resonance 
structure at large invariant mass. 
More specifically, the rate with gluon condensate is important 
for $M\le 200$ MeV whereas those from the $\rho-q$ interaction 
and HTLpt would be important for $M\ge 200$ MeV for the deconfined 
phase in heavy-ion collisions. Instead of considering only the Born-rate  
the various nonperturbative and perturbative  rates 
from appropriate domains of the invariant mass below $1$ GeV 
would comprise a more realistic rate for low mass dileptons from
the deconfined phase created in heavy-ion collisions. We hope that
more elaborate future lattice gauge theory studies on dileptons 
above the deconfined temperature can provide a more insight than  
present LQCD calculations on the low mass region, which could then verify 
the various model calculations on low mass dileptons above the
deconfined temperatures. Finally, we also have discussed a more 
realistic way to look into the quark-hadron duality hypothesis 
than it is advocated in the literature.

\begin{acknowledgments}
The authors are thankful to S. Sarkar for providing the result of
their calculations for in-medium hadron gas rate and P. Petreczky for 
also supplying the lattice data. MGM acknowledges various useful 
discussions and communications with H. van Hees. This work was 
partly supported by the Helmholtz International
Centre for FAIR within the framework of LOEWE (Landes-Offensive zur Entwicklung
Wissenschaftlich-{\"o}konomischer Exzellenz) program launched by the state of
Hesse, Germany.
\end{acknowledgments}


\begin{thebibliography}{99}

\bibitem{heinz} U. Heinz and M. Jacob, {\em `Evidence for a New State of 
Matter: An Assessment of the Result from CERN SPS Lead Beam Program'},
$\langle$ nucl-th/0002042 $\rangle$.

\bibitem{white} I. Arsene {\em et al.} (BRAHMS Collaboration), Nucl. Phys.
 {\bf A757}, 1 (2005); K. Adcox {\em et al.} (PHENIX Collaboration),
{\it ibid.} {\bf 757}, 184 (2005); B. B. Back {\em et al.}
(PHOBOS Collaboration), {\it ibid.} {\bf 757}, 28 (2005);
J. Adams {\em et al.} (STAR Collaboration), {\it ibid.} {\bf 757}, 102 (2005).

\bibitem{dilep} PHENIX Collaboration, A. Adare {\em et al.}, 
Phys. Rev. C {\bf 81}, 034911 (2010).

\bibitem{phot} PHENIX Collaboration, S. S. Adler {\it et al.},
Phys. Rev. Lett. {\bf 98}, 012002 (2007).

\bibitem{ellip} PHENIX Collaboration A. Adare {\it et al.},
Phys. Rev. Lett.{\bf 98}, 162301 (2007).

\bibitem{jet} PHENIX Collaboration, K. Adcox {\it et al.}, Phys. Rev.
Lett.{\bf 88}, 022301 (2002); STAR Collaboration, C. Adler {\it et al.},
Phys. Rev. Lett. {\bf 89}, 092302 (2002).

\bibitem{phenix} PHENIX Collaboration, S. S. Adler {\it et al.},
Phys. Rev. Lett. {\bf 91}, 172301 (2003); T. Chujo, PHENIX Collaboration,
Nucl. Phys. A{\bf 715}, 151c (2003).

\bibitem{larry} L. McLerran and T. Toimela, Phys. Rev. D{\bf 31}, 545 (1985).

\bibitem{agak} CERES Collaboration, G. Agakichiev {\em et al.}, 
Phys. Rev. Lett. {\bf 75},
1272 (1995); Phys. Lett. B{\bf 422}, 405 (1998); N. Masera for the 
HELIOS-3 Collaboration, Nucl. Phys. A {\bf 590}, 93c (1995); A Drees for
the CERES collaboration, Nucl. Phys. B{630}, 449c (1998).


\bibitem{rapp1} R. Rapp and J. Wambach, {\em `Chiral Symmetry Restoration
and Dileptons in relativistic Heavy-Ion Collisions'}, Adv. Nucl. Phys. 
{\bf 25}, 1 (2000).

\bibitem{rapp2} R. Rapp, J. Wambach, and H. van Hees, {\em `The Chiral
Restoration Transition of QCD and Low Mass Dileptons'}, arXive:0901.3289.

\bibitem{cass99} W. Cassing and E. L. Bratkovskaya, Phys. Rep. {\bf 308},
65 (1999).

\bibitem{sum} G. E. Brown and M. Rho, Phys. Rev. Lett.{\bf 66}, 2720 (1991);
B. Friman and H. J. Priner, Nucl. Phys. A{\bf 617}, 496 (1997); R. Rapp,
G. Chanfray, and J. Wambach, Nucl. Phys. A{\bf 617}, 472 (1997); 
Phys. Rev. Lett. {\bf 76}, 368 (1996); C. Gale and P. Lichard, Phys. Rev.
D{\bf 49}, 3338 (1994); R. Rapp and C. Gale, Phys. Rev. C{\bf 60}, 024903
(1999);  F. Klingl,
N. Kaiser, and W. Wiese, Nucl. Phys. A{\bf 624}, 527 (1997); W. Peters,
M. Post, H. Lenske, S. Leupold, and U. Mosel, Nucl. Phys. A{\bf 632}, 109
(1998); W. Cassing, E. L. Bratkovskaya, R. Rapp, and J. Wambach, Phys.
Rev. C{\bf 57}, 916 (1998); M. Post, S. Leupold, and U. Mosel, Nucl. Phys.
A {\bf 689}, 753 (2001); V. Koch, M. Bleicher, A. K. Dutt-Mazumder, C. Gale
and C. M. Ko, in {\em Hirschegg 2000, Hadrons in dense matter}, p.136;
D. K. Srivastava, B. Sinha, and C. Gale, Phys. Rev. C{\bf 53}, 567 (1996);
D. Pal, D. K. Srivastava, and K. Haglin, Phys. Rev. C{\bf 54}, 1366 (1996);
D. K. Srivastava, B. Sinha, D. Pal, C. Gale, and K. Haglin, Nucl. Phys. 
A{\bf 610}, 350c (1996); D. Pal and M. G. Mustafa, Phys. Rev. C{\bf 60},
034905 (1999); D. K. Srivastava, M. G. Mustafa, and B. M\"uller, Phys. Rev.
C{\bf 56}, 1064 (1997);
J. Alam, S. Sarkar, P. Roy, T. Hatsuda, and B. Sinha, Ann. Phys. {\bf 286},
159 (2001); J. Alam, P. Roy and S. Sarkar, Phys. Rev. C{\bf 67}, 054901
(2003). 

\bibitem{dz09} K. Dusling and I. Zahed, Nucl. Phys. A{\bf 825}, 212 (2009).

\bibitem{bcl09} E. L. Bratkovskaya, W. Cassing, and O. Linnyk, Phys. Lett.
B{\bf 670}, 428 (2009).

\bibitem{agkz98}P. Aurenche, F. Gelis, R. Kobes, and H. Zaraket, Phys. Rev.
D{\bf 58}, 085003 (1998).

%\bibitem{pert} P. Arnold and C. Zhai, Phys. Rev. D{\bf 50}, 7603 (1994);
%{\bf 51}, 1906 (1995); C. Zhai and Kastering, Phys. Rev. D{\bf 52}, 7232
%(1995); E. Braaten and A. Nieto, Phys. Rev. D {\bf 53}, 3421 (1996);
%Phys. Rev. Lett. {\bf 76}, 1417 (1996); 
%K. Kajantie, M. Laine, K. Rummukainen, and Y. Schroder, Phys. Rev. D{\bf 67}, 
%105008 (2006); A. Ipp, K. Kajantie, A. Rebhan, and A. Vourinen, 
%Phys. Rev. D{\bf 74}, 045016 (2006).

\bibitem{karsch} F. Karsch, E. Laermann, P. Petreczky, S. Stickan, and 
I. Wetzorke, Phys. Lett. B {\bf 530}, 147 (2000).

\bibitem{boyd} G. Boyd {\em et al.}, Nucl. Phys. B{\bf 469}, 419 (1996).

\bibitem{lateos} C. R. Alton {\em et al.}, Phys. Rev. D {\bf 68},
014507 (2003);C. R. Alton {\it et al.}, Phys. Rev. D {\bf 71}, 054508 (2005);
A. Bazavov {\it et al.}, Phys. Rev. D{\bf 80}, 014504
(2009); MILC Collaboration, C. Bernard {\it et al.}, Phys. Rev.
D{\bf 71}, 034504 (2005). 

\bibitem{peter} P. Petreczky, Nucl. Phys. A{\bf 830}, 11c (2009); 
P. Petreczky, arXive:1009.5935.

\bibitem{markus05} M. H. Thoma, J. Phys. G{\bf 31}, L7 (2005).

\bibitem{fair} $http://www.gsi.de/portrait/fair_-e.html$

\bibitem{qhdual} R. Rapp and J. Wambach, Eur. Phys. J. A{\bf 6}, 415 (1999).

\bibitem{cley} J. Cleymans, J. Fingberg, and K. Redlich, Phys. Rev. D{\bf 35}, 
2153 (1987).

\bibitem{gale} C. Gale and J. Kapusta, Nucl. Phys. B{\bf 357}, 65 (1991).

\bibitem{braaten} E. Braaten and R. D. Pisarski, Nucl. Phys. {\bf B337},
569 (1990); Phys. Rev. Lett. {\bf 64}, 1338 (1990).

\bibitem{agz00} P. Aurenche, F. Gelis, and H. Zaraket, Phys. Rev. D{\bf 61},
116001 (2000).

\bibitem{yaun} E. Braaten, R. D. Pisarski, and T. C. Yuan, Phys. Rev.
Lett. {\bf 64}, 2242 (1990).

\bibitem{markus} M. H. Thoma, Nucl. Phys. (Proc. Suppl.) B{\bf 92}, 162 (2001);
M. G. Mustafa and M. H. Thoma, Pramana {\bf 60}, 711 (2003); A. Peshier and
M. H. Thoma, Phys. Rev. Lett. {\bf 84}, 841 (2000).

\bibitem{munshi} F. Karsch, M. G. Mustafa, and M. H. Thoma, Phys. Lett. 
B{\bf 497}, 249 (2001). 

\bibitem{wong} S. M. H. Wong, Z. Phys. C{\bf 53}, 465 (1992).

\bibitem{agkz99}P. Aurenche, F. Gelis, R. Kobes, and H. Zaraket, 
Phys. Rev. D{\bf 60}, 076002 (1999).

\bibitem{tt97} M. H. Thoma and C. Traxler, Phys. Rev. D{\bf 56},
198 (1997).

\bibitem{cga08} M. E. Carrington, A. Gynther and P. Aurenche,
Phys. Rev. D{\bf 77}, 045035 (2008); P. Aurenche, F. Gelis, G. D. Moore,
and H. Zaraket, J. High. Energy Phys. {\bf 12}, 006 (2002); 
J. High. Energy Phys. {\bf 07}, 063 (2002). 

\bibitem{ar92} T. Altherr and P. V. Ruuskanen, Nucl. Phys. B{\bf 380}, 
377 (1992). 

\bibitem{cdj94} J. Cleymans, I. Dadic, and J. Joubert, Phys. Rev. 
D{\bf 49}, 230 (1994); J. Cleymans and I. Dadic, Phys. Rev. D{\bf 47},
160 (1993). 

\bibitem{kls92} J. I. Kapusta, P. Lichard, and D. Seibert, Phys. Rev. D
{\bf 44}, 2774 (1992); R. Baier, H. Nakkagawa, A. Niegawa, and K. Redlich,
Z. Phys. C{\bf 53}, 433 (1992).
 
\bibitem{aa89}T. Altherr and P. Aurench, Z. Phys. C{\bf 45}, 99 (1989).
 
\bibitem{kw00} J. I. Kapusta and S. M. H. Wong, Phys. Rev. C{\bf 62},
027901 (2000).

\bibitem{abb00} P. Aurenche {\em et al.}, Phys. Rev. D{\bf 65}, 038501
(2002).

\bibitem{pkps94} A. Peshier, B. K\"ampfer, O. P. Pavlenko, and G. Soff,
Phys. Lett. B {\bf 337}, 235 (1994); Phys. Rev. D{\bf 54}, 2399 (1996);
U. Heinz and P. Levai, Phys. Rev. C{\bf 57}, 1987 (1998).

\bibitem{st99} A. Sch\"afer and M. H. Thoma, Phys. Lett. B{451}, 195
(1999).

\bibitem{mst00} M. G. Mustafa, A. Sch\"afer and M. H. Thoma, 
Phys. Lett. B{\bf 472}, 402 (2000).

\bibitem{mst99} M. G. Mustafa, A. Sch\"afer and M. H. Thoma, 
Phys. Rev. C{\bf 61}, 024902 (1999); Nucl. Phys. A{661}, 653 (1999).

\bibitem{tlm00} M. H. Thoma, S. Leupold, and U. Mosel, Eur. Phys. J A{\bf 7},
219 (2000)

\bibitem{clov} B. Sheikholeslami and R. Wohlert, Nucl. Phys. B{\bf 259},
572 (1985); M. L\"uster {\em et al.}, Nucl. Phys. B{\bf 491}, 344 (1997).

\bibitem{mem} Y. Nakahara, M. Asakawa, and T. Hatsuda, Phys. Rev. D{\bf 60},
091503 (1999); M. Asakawa, T. Hatsuda, and Y. Nakahara, Prog. Part. Nucl. Phys.
{\bf 46}, 459 (2001); I. Wetzorke, F. Karsch, in: C. P. Korthals-Altes (Ed.),
Proceedings of the International Workshop on Strong and Electroweak Matter,
World Scientific, 2001, p.193.

\bibitem{gsa10} S. Ghosh, S. Sarkar, and J. Alam, arXiv:1009.1260.

\bibitem{private} H. van Hees: Private communication

\end{thebibliography}
\end{document}